\documentclass[11pt,a4paper]{article}
\usepackage{jheppub}
\usepackage{subcaption}
\usepackage{ulem}
\usepackage{graphicx}% Include figure files
\usepackage{bm}% bold math
\usepackage{amsmath}
\usepackage{mathrsfs}
\usepackage{multirow}
\usepackage{xcolor}
\usepackage{enumitem}

\newcommand{\gsim}{\gtrsim}
\newcommand{\lsim}{\lesssim}

\newcommand{\lf}{\left(}
\newcommand{\ri}{\right)}

\newcommand{\nn}{\nonumber}

\newcommand{\sqt}{\sqrt{2}}
\newcommand{\hf}{\frac{1}{2}}

\renewcommand{\lg}{\mathscr{L}}
\newcommand{\mco}{\mathcal{O}}

\newcommand{\br}{\mathcal{B}}
\newcommand{\hc}{{\rm H.c.}}

\newcommand{\sm}{{\rm SM}}

\newcommand{\nb}{{\;{\rm nb}}}
\newcommand{\pb}{{\;{\rm pb}}}
\newcommand{\fb}{{\;{\rm fb}}}
\newcommand{\ifb}{{\;{\rm fb}^{-1}}}
\newcommand{\iab}{{\;{\rm ab}^{-1}}}

\newcommand{\kev}{{\;\,{\rm keV}}}

\newcommand{\gev}{{\;{\rm GeV}}}
\newcommand{\tev}{{\;{\rm TeV}}}

\newcommand{\beq}{\begin{equation}}
\newcommand{\eeq}{\end{equation}}
\newcommand{\bea}{\begin{eqnarray}}
\newcommand{\eea}{\end{eqnarray}}
\newcommand{\barr}{\begin{array}}
\newcommand{\earr}{\end{array}}
\newcommand{\bc}{\begin{center}}
\newcommand{\ec}{\end{center}}
\newcommand{\bit}{\begin{itemize}}
\newcommand{\eit}{\end{itemize}}
\newcommand{\ben}{\begin{enumerate}}
\newcommand{\een}{\end{enumerate}}
\newcommand{\dt}{\delta}
\newcommand{\Dt}{\Delta}

\newcommand{\sg}{\sigma}

\newcommand{\kp}{\kappa}

\newcommand{\gm}{\gamma}
\newcommand{\Gm}{\Gamma}
\newcommand{\lm}{\lambda}

\newcommand{\mch}{M_{H^\pm}}

\renewcommand{\mho}{m_{125}}
\newcommand{\mh}{m_{H}}

\newcommand{\ee}      {{e^+ e^-}}

\newcommand{\ttop}      {{t\bar{t}}}

\newcommand{\ww}      {{W^+_\mu W^{-\mu}}}
\newcommand{\zz}      {{Z_\mu Z^\mu}}
\renewcommand{\aa}      {{A_\mu A^\mu}}

\newcommand{\qq}      {{q \bar{q}}}
\newcommand{\chh}      {{H^+ H^-}}
\newcommand{\met}      {{E_T^{\rm miss}}}

\newcommand{\neuo}{{\tilde{\chi}^0_1}}
\newcommand{\chao}{{\tilde{\chi}^\pm_1}}

\renewcommand{\mho}{M_{h_1}}
\newcommand{\mht}{M_{h_2}}

\begin{document}

\vspace*{18pt}

\begin{flushright}
{}
\end{flushright}

\title{Phenomenology of the Inert Doublet Model \\
with a global $\mathbf{U(1)}$ symmetry}
\author[a]{Adil Jueid,}
\author[a]{Jinheung Kim,}
\author[a]{Soojin Lee,}
\author[b]{So Young Shim,}
\author[a]{and Jeonghyeon Song}

\affiliation[a]{Department of Physics, Konkuk University, Seoul 05029, Korea}
\affiliation[b]{Center for Theoretical Physics of the Universe, Institute for Basic Science (IBS), Daejeon, 34126, Korea}

\emailAdd{adil.hep@gmail.com}
\emailAdd{jinheung.kim1216@gmail.com}
\emailAdd{soojinlee957@gmail.com}
\emailAdd{soyoung@ibs.re.kr}
\emailAdd{jhsong@konkuk.ac.kr}

%\begin{abstract}
\abstract{
The inert doublet model is a minimal dark matter model
with strong theoretical motivations,
where the stability of dark matter is usually achieved by imposing a $Z_2$ parity.
We promote the $Z_2$ parity into a global $U(1)$ symmetry
and study its phenomenological implications.
There are two characteristic features of the model:
both the \textit{CP}-even and \textit{CP}-odd neutral inert scalars, 
$h_1$ and $h_2$, become DM candidates;
the number of model parameters is one less than that with $Z_2$ parity.
We first analyze the constraints from LEP experiments, electroweak precision tests,
theoretical stability, Higgs precision data, dark matter relic density,
and direct detection experiments.
It is found that if the model is required to explain at least 10\% of the observed relic density,
the theory is extremely limited such that
the dark matter mass is about $70\;{\rm GeV}$ and 
the charged Higgs boson is not very heavy. 
Focusing on this narrow parameter space,
we calculate the production cross sections of almost all the possible mono-$X$ and mono-$XX'$ processes
at the LHC. The mono-$W\gamma$ process is shown to have high discovery potential
with the help of the decay of the intermediate-mass charged Higgs boson into $W^\pm h_{1,2}$.
A search strategy is designed to increase the potential discovery of the model 
for the mono-$W\gamma$ signal at both the HL-LHC and the FCC-hh.
The optimal cut on  $E_T^{\rm miss}/\sqrt{H_T}$
is suggested to maximize the signal significance,
being about $0.76$ at the HL-LHC and
about $7.5$ at the FCC-hh.
}
%\end{abstract}

\maketitle

\section{Introduction}

One of the most convincing pieces of evidence that the standard model (SM) 
is not the final theory of particle physics is the observed dark matter (DM)
in the Universe~\cite{Bertone:2004pz}.
Not knowing what it is yet,
DM searches have been enthusiastically performed in three directions,
direct detection,
indirect detection,
and its production at high energy colliders.
Undeniably, a theory is essential in understanding all the different experimental results
as a whole.
A convenient approach to the theory of DM is
through high-dimensional operators in the effective field theory~\cite{Cao:2009uw,Beltran:2010ww,Bai:2010hh,Goodman:2010qn,Fox:2011pm,Busoni:2013lha,Beniwal:2015sdl,Bishara:2016hek}
or through simplified DM models~\cite{An:2012va,Frandsen:2012rk,Dreiner:2013vla,Abdallah:2015ter,Boveia:2016mrp,Kahlhoefer:2015bea,Bauer:2017ota,DeSimone:2016fbz}.
Nevertheless,
studying one complete DM model 
has enormous advantages, especially for the DM searches at a high energy collider.
For example, if the theory accommodates other new heavy particles 
decaying into a DM particle and an SM particle,
some mono-$XX'$ processes can be as important as mono-$X$ processes.
We may have a different golden mode.
In addition,
a complete theory makes it possible to require theoretical stability
and the compatibility with the electroweak precision data.
The comprehensive study of the whole constraints including the DM observations
shall significantly limit the viable region of the parameter space of the model.% especially 
%when the number of model parameters is not too many.
When the allowed parameter space is narrow enough,
we could predict more definite signatures at a high energy collider.
The inevitable weakness, the model-dependence, is something 
we can overcome only by dedicated studies on the phenomenology of  each viable DM model.

One good example of theoretically well-motivated DM models 
with a relatively small number of parameters
is the inert doublet model (IDM) with $Z_2$ parity~\cite{Deshpande:1977rw}.
As one of the simplest extensions of the SM, 
the IDM introduces an additional Higgs doublet field $\Phi'$
which is odd under the $Z_2$ parity transformation.
The model has drawn a lot of interest due to its capabilities 
such as triggering the first order electroweak phase transition~\cite{Chowdhury:2011ga,Borah:2012pu,Blinov:2015vma,Cline:2013bln}, and
generating neutrino masses~\cite{Ma:2006km}.
Most of all, $\Phi'$ that has neither a vaccum expectation value nor couplings to the SM fermions
provides good DM candidates, neutral inert scalar bosons~\cite{LopezHonorez:2006gr,Hambye:2009pw,LopezHonorez:2010tb,Goudelis:2013uca,Dolle:2009fn}.
In the literature, various phenomenological implications of the model have been extensively studied~\cite{Gustafsson:2007pc,Cao:2007rm,Agrawal:2008xz,Andreas:2009hj,Dolle:2009ft,Nezri:2009jd,Miao:2010rg,Gustafsson:2012aj,Arhrib:2012ia,Swiezewska:2012eh,Krawczyk:2013jta,Garcia-Cely:2013zga,Arhrib:2013ela,Ilnicka:2015ova,Belanger:2015kga,Ilnicka:2015jba,Belyaev:2016lok,Poulose:2016lvz,Belyaev:2018ext}.

Further simplification of the IDM was made 
by introducing a Peccei-Quinn symmetry to protect 
the mass degeneracy between the lightest and next-to-lightest neutral scalars
for the inelastic DM-nucleus scattering~\cite{Arina:2009um}.
Recently, the idea was extended
to explore the mass degeneracies among some of the scalar bosons 
in various multi Higgs doublet models~\cite{Haber:2018iwr}.
Focusing on the IDM,
we find that promoting the $Z_2$ parity into a global $U(1)$ symmetry has two immediate consequences:
(i) two neutral inert scalar bosons become DM particles;
(ii) the number of model parameters is one less than that of the IDM with $Z_2$ parity.
We naturally expect that the parameter space will be very strongly
restricted by theoretical and experimental constraints, which
should be investigated by a comprehensive study on the phenomenology of the model.
Thus, our first purpose in this paper is to perform a comprehensive analysis of this model.
%This is our first purpose.
With the result of the allowed (possibly very small) parameter space,
we can assess all possible mono-$X$ and mono-$XX'$ processes
at the high-luminosity LHC
(HL-LHC) at $\sqrt{s}=14\tev$ with $3\iab$~\cite{Apollinari:2017cqg}
and the FCC-hh at $\sqrt{s}=100\tev$ and $30\iab$ \cite{Benedikt:2018csr},
and suggest a golden mode for this model. 
%It is the main result.

The paper is organized in the following way.
In Sec.~\ref{sec:review},
we briefly review the IDM with a global $U(1)$ symmetry.
Section \ref{sec:constraints} deals with various constraints
such as LEP experiments, the electroweak oblique parameters,
the stability of scalar potential, unitarity, 
Higgs precision data including the Higgs invisible decay rate and $\kp_\gm$,
DM relic density, and direct detection experiments.
In Sec.~\ref{sec:LHC},
we calculate the total production cross sections of
major mono-$X$ and mono-$XX'$ processes at the HL-LHC.
Projecting the current direct DM searches onto the HL-LHC,
we will suggest that the mono-$W\gm$ is one of the most efficient channels 
to probe the model.
In section \ref{sec:simulation}, we present a search strategy to look for the model in the mono-$W\gm$ process at the HL-LHC and FCC-hh. We conclude in Sec.~\ref{sec:conclusion}.

%%%%%%%%%%%%%%%%%%%%%%%%%%%%%%%%%%%%
\section{Brief review of the IDM with a continuous $U(1)$ symmetry}
\label{sec:review}
%%%%%%%%%%%%%%%%%%%%%%%%%%%%%%%%%%%%
In the IDM, the scalar sector is augmented by one extra $SU(2)_L$ doublet
$\Phi'$, in addition to the SM one $\Phi$.
And we introduce a global $U(1)$ symmetry, 
under which $\Phi$ and $\Phi'$ transform as
\bea
\label{eq:Phi:def}
\Phi \to \Phi, \quad \Phi' \to e^{i \theta} \Phi'.
\eea
%All of the other SM fields have zero $U(1)$ charge.
The fact that the extra doublet $\Phi'$ has a non-zero $U(1)$ charge implies that 
its vacuum expectation value is vanishing. We have
\bea
\label{eq:inert:vacuum}
\langle \Phi_0 \rangle=\frac{v}{\sqrt{2}}\approx 174\gev,
\quad
\langle \Phi^{\prime}_{ 0}\rangle  \equiv \frac{v_D}{\sqrt{2}  } =0.
\eea
In the unitary gauge, $\Phi$ and $\Phi'$ are written as
\bea
\label{eq:Phi:def}
\Phi = 
\lf
\barr{c}
0 \\ \frac{1}{\sqt}(v+ H)
\earr
\ri,
\quad
\Phi' = 
\lf
\barr{c}
H^+ \\ \frac{1}{\sqt}(h_1 + i h_2)
\earr
\ri.
\eea
The particle spectrum of the model consists of five scalar states, 
the SM Higgs boson $H$ and the inert scalar bosons $h_1$, $h_2$ and $H^\pm$.
%which yield five physical Higgs bosons, 
%the SM-like Higgs boson $H$ and the inert scalar bosons of $h_1$, $h_2$, and $H^\pm$. 
Although $h_1$ and $h_2$ have opposite \textit{CP} parities, 
we cannot tell which one is which,
because the \textit{CP} transformation properties of $h_1$ and $h_2$ are exchanged 
under re-phasing of $\Phi' \to i \Phi'$~\cite{Belyaev:2016lok}.
The most general renormalizable and \textit{CP} invariant scalar potential that preserves the additional $U(1)$ symmetry is given by
\bea
\label{eq:potential}
V(\Phi,\Phi') &=&
- \mu_1^2 \Phi^\dagger \Phi + \mu_2^2 \Phi^{\prime \dagger} \Phi'
+ \lm_1  \lf \Phi^\dagger \Phi \ri^2 + \lm_2 \lf  \Phi^{\prime \dagger} \Phi' \ri^2
\\ \nn &&
+\lm_3  \lf \Phi^\dagger \Phi \ri  \lf  \Phi^{\prime \dagger} \Phi' \ri
+ \lm_4 \lf \Phi^\dagger  \Phi' \ri \lf   \Phi^{\prime \dagger} \Phi \ri 
,
\eea
where $\mu_1^2>0$ and $\mu_2^2>0$. 
The usual $\lm_5$ term, 
proportional to $\big\{\lf \Phi^\dagger\Phi'  \ri^2 +\hc \big\}$,
is prohibited by the $U(1)$ symmetry. Therefore, the absence of this term 
leads to the mass degeneracy between $h_1$ and $h_2$: 
\bea
\mho = \mht   \equiv M_S.
\eea
In what follows, we will call this model the IDM-$U(1)$ to distinguish it from %, in distinction from
the ordinary IDM model with $Z_2$ parity. 

The IDM-$U(1)$ contains six additional parameters,  
$\mu_1^2$, $\mu_2^2$, and $\lm_{1,2,3,4}$. 
Since the Higgs boson mass fixes %determines 
two parameters $\mu_1^2$ and $\lm_1$ as
\bea
\label{eq:m1:lm1}
\mu_1^2 &=& \frac{\mh^2}{2},\quad \lm_1 = \frac{\mh^2}{2 v^2},
\eea
we are left with only %there are 
four extra parameters. % left.
We take the physical parameter basis defined by
\bea
\{
M_S,\mch,\lm_{L},\lm_2
\},
\eea
where $\lm_L = (\lm_3+\lm_4)/2$.
The other model parameters are obtained from the following relations:
\bea
\label{eq:parameters}
 \mu_2^2 &=& M_S^2 - \lm_{L} v^2,
\\ \nn
\lm_3 &=& 2 \left[ \lm_{L}+\frac{\mch^2 -M_S^2}{v^2} \right],
\\ \nn
\lm_4   &=& -\frac{2}{v^2}
\lf
\mch^2-M_S^2
\ri .
\eea

The interaction Lagrangian of the SM Higgs boson $H$ with the SM particles
is the same as in the SM. The gauge interaction Lagrangian of the inert scalar bosons is
\bea
\label{eq:Lg:gauge}
\lg_{\rm gauge} &=&
\hf g_Z Z^\mu h_2 \stackrel{\leftrightarrow}{\partial}_\mu \!\! h_1
-\frac{1}{2}g 
\left[
iW_\mu^+ H^- \stackrel{\leftrightarrow}{\partial}_\mu \!\! (h_1+i h_2) +\hc
\right]
\\ \nn &&
+ i \left[
 e A^\mu +  g_{H^\pm}  Z^\mu
\right]H^+ \stackrel{\leftrightarrow}{\partial}_\mu \!\! H^-
%\\ \nn &&
+
\lf
\frac{1}{4} g^2 \ww + \frac{1}{8} g_Z^2 \zz 
\ri (h_1^2+ h_2^2)
\\ \nn &&
+
\left[
\hf g^2 \ww + e^2 \aa + g_{H^\pm}^2 \zz + 2 e g_{H^\pm} A_\mu Z^\mu
\right] H^+ H^-
\\[3pt] \nn &&
+ \left[
\lf \hf e g A^\mu W^{+}_{\mu} - \hf g_Z g s_W^2 Z^\mu W^+_\mu \ri H^-(h_1+i h_2) 
+ \hc
\right],
\eea
where $g_{ H^\pm} = g_Z \left(1/2 -s_W^2 \right)$, $g_Z = g/c_W$,
$s_W=\sin\theta_W$, $c_W=\cos\theta_W$, and $\theta_W$ is the electroweak mixing angle.
The interactions of the inert scalar bosons to a single $H$ are described by 
\bea
\label{eq:Lg:scalar}
\lg_{\rm scalar} & \supset &  - v H 
\left[
\lm_L (h_1^2+ h_2^2) + \lm_3 \chh
\right]
.
%\\[3pt] \nn
% && -\frac{1}{2} \lm_{L} H^2 (h_1^2+ h_2^2) 
% -\hf \lm_3 H^2 \chh.
% \\ \nn &&
% - \lm_2 
% \left[
% \frac{1}{4} \lf h_1^2+ h_2^2 \ri^2 + \chh \lf h_1^2+ h_2^2 + \chh \ri^2
% \right].
\eea
Since the SM fields do not have the $U(1)$ charge, 
the Yukawa couplings of the inert scalar bosons to the SM fermions vanish.
Consequently, the decays of the new scalar bosons are very simple.
Both $h_1$ and $h_2$, as the lightest particles with nonzero $U(1)$ charge,
do not decay and become the DM candidates.
The charged Higgs boson $H^\pm$ exclusively decays into $W^{\pm (*)} h_{1,2}$:
\bea
\br(H^\pm \to W^{\pm (*)} h_{1,2}) = 1.
\eea

Brief comments on the necessity for the soft breaking of the global $U(1)$
symmetry are in order here.
The $U(1)$ symmetry protects the exact mass degeneracy 
between $h_1$ and $h_2$.
Then the $Z$-$h_1$-$h_2$ vertex in Eq.~(\ref{eq:Lg:gauge})
causes large inelastic DM-nucleus scattering
in the direct detection experiments,
which is excluded by the current results~\cite{TuckerSmith:2001hy,Arina:2009um}.
If we allow very small mass difference like $\dt m (\equiv \mht -\mho)\gsim 200 \kev$,
the inelastic scattering does not occur
because of the kinematical threshold 
for the DM to scatter inelastically off a nucleus.\footnote{
There is
another interesting possibility of inelastic scattering of the DM 
in simultaneously explaining the annual modulation measured
by DAMA~\cite{Bernabei:2008yi,Bernabei:2010mq} and the null results
of the other DM direct detection experiments by tuning 
the mass difference as $\dt m \approx 13\kev$~\cite{Cui:2009xq,TuckerSmith:2004jv,TuckerSmith:2001hy,Chang:2008gd,Petriello:2008jj}.}
With very small $\dt m $,
the phenomenological signatures as well as 
the theoretical stability, the Higgs precision constraints,
and the cosmological relic density
are practically the same as in the IDM with exact $U(1)$ symmetry.
We can attribute the soft-breaking
%Attributing the soft breaking 
of the $U(1)$ symmetry to some high dimensional operators
from an unknown UV theory.
Therefore, we adopt the IDM-$U(1)$ and focus on its phenomenological study.

%%%%%%%%%%%%%%%%%%%%%%%%%%%%
\section{Theoretical and experimental constraints} 
\label{sec:constraints}
%%%%%%%%%%%%%%%%%%%%%%%%%%%%
Compared to the IDM with discrete $Z_2$ parity, 
the IDM-$U(1)$ is severely restricted due to the compressed spectrum in the neutral component of the scalar doublet $\Phi'$.
In this section, we consider the followings constraints:
\bit
\item $Z$-decay width and the bounds from the searches of charginos at LEP;
\item The Electroweak Precision Data (EWPD) encoded in the Peskin-Takeuchi oblique parameters;
\item Theoretical constraints from boundness-from-below (BFB) of the scalar potential, 
perturbativity, and unitarity;
\item Searches of Higgs boson invisible decays at the LHC and the precision measurement of the Higgs coupling modifier to a photon pair, $\kp_\gm$;
\item DM relic density and the bounds from DM direct detection experiments.% direct detection experiments.
\eit
The bounds from the direct search of a pseudo-scalar boson $A$
reported by the LEP, Tevatron, and LHC 
are not relevant in this model
since all of these searches depend on the fermionic decay modes of $A$ while, in our model, %$ \to \bar{f} f$.
the pseudo-scalar state (either $h_1$ or $h_2$) does not decay.

\subsection{LEP experiments and electroweak precision data}

In the IDM-$U(1)$, the $Z$ boson decays into $h_1 h_2$ if kinematically allowed: see Eq.~(\ref{eq:Lg:gauge}).
The precise measurement of the total width of the $Z$ boson
excludes $M_S < m_Z/2$~\cite{Tanabashi:2018oca}.
In addition, the reinterpretation of the chargino pair production
as $\ee\to H^+ H^-$~\cite{Pierce:2007ut, Blinov:2015qva}
puts a lower bound on the charged Higgs boson mass.
In summary, the null results from LEP require 
\bea
\label{eq:LEP}
M_S > \frac{m_Z}{2}, \quad \mch > 70\gev.
\eea

One of the most significant constraints on the IDM-$U(1)$ is from the
EWPD oblique parameters, $S$ and $T$.
The contributions of the inert scalar bosons to the $S$ and $T$ parameters
can be written as 
\bea
S &=& \frac{1}{12\pi}
 \ln \frac{M_S^2}{\mch^2} ,
\\ \nn
T &=& \frac{1}{16\pi^2 \alpha v^2} F(\mch^2,M_S^2),
\eea
where
the loop function $F(x,y)$ is 
\bea
F(x,y) =
\left\{
\begin{array}{ll}
\frac{x+y}{2}-\frac{xy}{x-y}\ln \frac{x}{y}, & \hbox{ if } x \neq y;
\\
0, & \hbox{ if } x =y.
\end{array}
\right.
\eea 
The current best-fit results are given by ~\cite{Tanabashi:2018oca}
\bea
S &=& 0.02 \pm 0.07,
\quad
T = 0.07 \pm 0.06, \quad
 \rho_{ST}=0.92,
\eea
where $U=0$ is assumed, and with $\rho_{ST}$ is the correlation between $S$ and $T$.
In order to obtain the allowed region in the mass spectrum of the model, we minimize the following $\chi^2$:
%In order to include the strong correlation $\rho_{ST}$, we minimize the following $\chi^2$:
\bea
\chi^2= \sum_{\mco=S,T} \frac{(\mco-\mco_{\rm exp})^2}{\sg_\mco^2 (1-\rho_{ST})}
- 2 \rho_{ST}\frac{(S-S_{\rm exp})(T-T_{\rm exp})}{\sg_S \sg_T (1-\rho_{ST})},
\eea
where $\sg_\mco$ is the error on the observable $\mco$.

\begin{figure}[h] \centering
\begin{center}
\includegraphics[width=.618\textwidth]{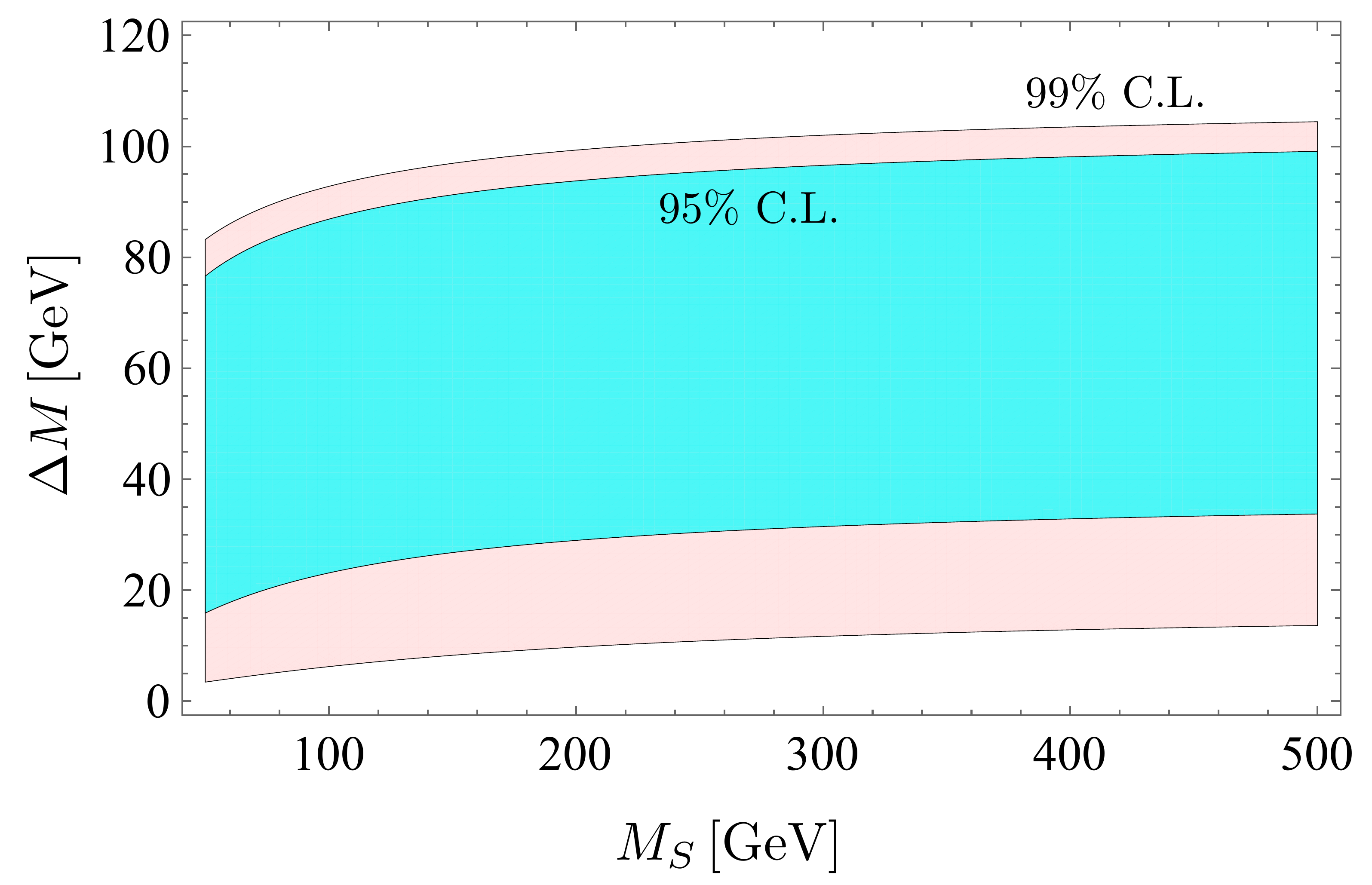}
\caption{\label{fig-STconstraint}
Allowed region of $(M_S,\Dt M)$ by the electroweak oblique parameters $S$ and $T$ at 95\% and 99\% C.L., where $M_S \equiv \mho=\mht$ and $ \Dt M \equiv \mch - M_S$.
The strong correlation between $S$ and $T$, $\rho_{ST}=0.92$,
is included in the $\chi^2$ calculation.
}
\end{center}
\end{figure}

In Fig.~\ref{fig-STconstraint}, we present 
the allowed region of $(M_S,\Dt M)$ 
by the electroweak oblique parameters $S$ and $T$ at 95\% and 99\% C.L.,
where $\Dt M \equiv \mch - M_S$. 
We can see that the constraints from EWPD on the mass spectrum are very stringent. 
In fact, the mass difference between 
the charged Higgs boson and the DM 
cannot be too large or too small, $10\lsim \Dt M \lsim 100\gev$.
The upper bound on $\Dt M$ implies that 
we cannot arbitrarily ignore $H^\pm$ by decoupling it from the model. 
%At the LHC, the  production of $h_{1,2}$ via $H^\pm \to W^{\pm(*)} h_{1,2}$ can be significant.
On the other hand, the presence of the lower bound on $\Dt M$ 
excludes the possibility of the total mass degeneracy, $\mho=\mht\simeq \mch$.\footnote{If $\Dt M \simeq 0.2\gev$, we could observe the disappearing charged track signatures at the LHC from the $H^\pm$ production and its subsequent soft decays~\cite{CMS:2014gxa}.}
Note that the correlation $\rho_{ST}$
plays a crucial role here: the allowed range would be 
$0 \leq \Dt M \lsim 190\gev$ if we assumed that $S$ and $T$ are uncorrelated.
%without it,
%the allowed range is $0 \leq \Dt M \lsim 190\gev$.
%In what follows, we consider two benchmark points, $\Dt M=85\gev$
%and $\Dt M =50\gev$.
%The $\Dt M=85\gev$ case allows the on-shell decay of $H^\pm \to W^\pm h_{1,2}$
%while the $\Dt M =50\gev$ case yields $H^\pm \to W^{\pm *} h_{1,2}$.

%%%%%%%%%%%%%%%%%%%%%%%%%%%%%%%%%%%%%%%%%
\subsection{Constraints from the theoretical stability and the Higgs precision data}
%%%%%%%%%%%%%%%%%%%%%%%%%%%%%%%%%%%%%%%%%

The parameters of the scalar potential should satisfy the following conditions
for the feasibility of the model.
\ben
\item Perturbativity: 
\bea
|\lm_{1,2,3,4}|\leq 8 \pi.
\eea
\item BFB condition for the scalar potential:
\bea
\label{eq:bounded}
\lm_{2}>0,\quad \lm_3>-2 \sqrt{\lm_1\lm_2}, \quad 
\lm_{L}>-\sqrt{\lm_1\lm_2}.
\eea
\item Electro-neutrality of the vacuum and the DM particle:
\bea
\lm_4 < 0.
\eea
\item Tree level unitarity:
the eigenvalues of the $S$ matrix for the scalar-scalar scattering processes
should satisfy~\cite{Arhrib:2012ia}
\bea
|a_{i}|\leq 8 \pi, \quad (i=1,2,\cdots,10)
\eea
where %$i=1,2,\cdots,10$ and
\bea
a_{1,2} &=& \lm_3 \pm \lm_4, \quad a_3 = \lm_3, \quad 
a_4=\lm_3+2 \lm_4,
\\ \nn
a_{5,6} &=& -\lm_1-\lm_2 \pm\sqrt{ (\lm_1-\lm_2)^2+\lm_4^2},
\\ \nn
a_{7,8} &=& -3\lm_1-3\lm_2 \pm \sqrt{9(\lm_1-\lm_2)^2+(2\lm_3+\lm_4)^2},
\\ \nn
a_{9,10} &=& - \lm_1 -\lm_2 \pm |\lm_1 - \lm_2|.
\eea
\item The invisible decay of the Higgs boson. 
If $M_S < m_H /2$, the SM-like Higgs boson decays invisibly via  
$H \to h_1 h_1/ h_2 h_2$.
The upper bound on the branching ratio of the invisible Higgs decay constrains 
$\lm_{L}$ as
\bea
\label{eq:lm34:constraint}
|\lm_{L}| < 
\left[
\frac{  \pi g^2\mh  \Gm_H^\sm }
{\beta_S m_W^2 \lf \frac{1}{\br_{\rm inv}^{\rm max} }-1 \ri  }
\right]^{1/2},
\eea 
where $\beta_S = \sqrt{1-4 M_S^2/m_H^2}$.
In our model with \textit{two} DM particles,
the upper bound on $|\lm_{L}|$ is smaller by a factor of $1/\sqrt{2}$ 
than that with one DM particle~\cite{Belyaev:2016lok}.
We adopt the latest ATLAS Higgs combined results~\cite{Aad:2019mbh},
$\br_{\rm inv}< 0.30$ at  95\% C.L.
\item The diphoton decay rate of the Higgs boson,
which is modified by the contributions from the charged Higgs boson.
The Higgs coupling modifier $\kp_\gm$ in the IDM-$U(1)$
is~\cite{Arhrib:2012ia,Swiezewska:2012eh,Krawczyk:2013jta}
\bea
\label{eq:Gm:Hrr}
\kp_\gm &=& 
\left|
\frac{
\sum_{i=t,b,\tau} N^C_{i} Q_i^2 A_{1/2}(\tau^h_f) + A_1(\tau^h_W)+\frac{\lm_3 v^2}{2 \mch^2} A_0(\tau^h_{H^\pm})}{
\sum_{i=t,b,\tau} N^C_{i} Q_i^2 A_{1/2}(\tau^h_f) + A_1(\tau^h_W)}
\right|,
\eea
where $N^C_{i}$ is the color factor of the fermion,
$\tau^h_i = {m_H^2}/(4 m_i^2)$,
and loop functions $A_{0,1/2,1}(\tau)$ can be found in e.g.~\cite{Krawczyk:2013jta}.
We take $\left|
\kp_\gm 
\right| = 1.05 \pm 0.09$~\cite{Aad:2019mbh}.
\een

Before presenting the results of the constraints,
we show that the BFB condition makes the inert vacuum in Eq.~(\ref{eq:inert:vacuum})
be the only true vacuum in the IDM-$U(1)$.
The scalar potential in Eq.~(\ref{eq:potential})
has an additional minimum, called the mixed extremum~\cite{Ginzburg:2010wa},
having the vacuum expectation value of $\Phi'$ as
\bea
\label{eq:vdsq}
\lf v_D^{\rm mix} \ri^2 = 
-\frac{\lm_1 \mu_2^2 + \lm_{L} \mu_1^2}{\lm_1 \lm_2 (1-R^2)},
\eea 
where $R \equiv \lm_{L}/\sqrt{\lm_1\lm_2}$.
The energy difference between the inert and mixed vacua is
\bea
\label{eq:DE}
\mathcal{E}_{\rm inert} -
\mathcal{E}_{\rm mixed}   = 
\frac{ \lf \lm_{L} \mu_1^2 + \lm_1 \mu_2^2 \ri^2}{4 \lm_1^2 \lm_2 (1-R^2)}.
\eea
If $\mathcal{E}_{\rm inert} -
\mathcal{E}_{\rm mixed} <0$, 
the inert vacuum becomes the global minimum.
The BFB condition in Eq.~(\ref{eq:bounded})
leads to $R > -1$.
When $1<R^2$, $\mathcal{E}_{\rm inert} <
\mathcal{E}_{\rm mixed}$ is automatically satisfied. 
If $R^2<1$, $\lf v_D^{\rm mix} \ri^2$ with Eqs.~(\ref{eq:m1:lm1}) and (\ref{eq:parameters})
becomes
\bea
\lf v_D^{\rm mix} \ri^2 =
-\frac{m_H^2}{2 v^2}
\frac{M_S^2}{\lm_1\lm_2 (1-R^2)}<0,
\eea
which disqualifies the mixed vacuum as a vacuum solution.
In summary, the IDM-$U(1)$ accommodates only one true vacuum, the inert vacuum.

\begin{figure}[h] \centering
\begin{center}
\includegraphics[width=.7\textwidth]{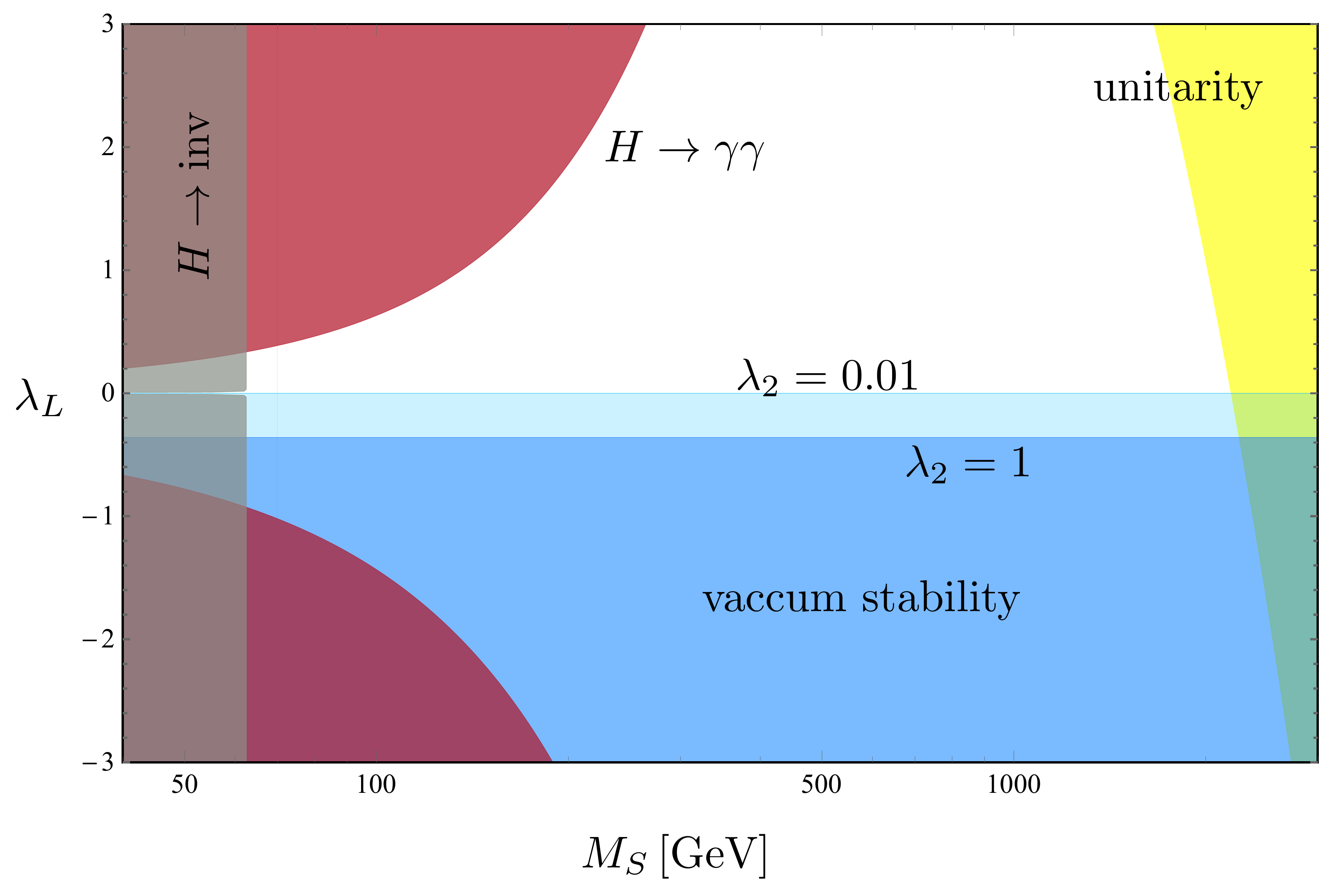}
\caption{\label{fig-theoryconst-85}
Excluded region of $(M_S,\lm_{L})$ by the unitarity,
the bounded-from-below scalar potential,
the diphoton decay rate of the Higgs boson, and the Higgs invisible decay rate.
We take $\mch =M_S + 85\gev$ for 
two cases of $\lm_2=0.01$ and $\lm_2=1$.
}
\end{center}
\end{figure}

In Fig.~\ref{fig-theoryconst-85},
we present the excluded regions of $(M_S,\lm_{L})$ by the unitarity,
the BFB scalar potential,
the Higgs coupling modifier $\kp_\gm$, and 
the Higgs invisible decay rate.
The perturbativity imposes weaker constraints than the unitarity
which do not appear in the figure.
We take  $\Dt M= 85\gev$ for two cases of $\lm_2=0.01$ and $\lm_2=1$.
Note that the exclusions barely depend on the mass splitting $\Dt M$:
if we set $\Dt M = 40\gev$, nothing will practically change.
The unitarity condition excludes heavy DM masses, $M_S$,
since the scalar quartic couplings $\lm_3$ and $\lm_4$
are proportional to $M_S$ for a given $\Dt M$: see Eq.~(\ref{eq:parameters}).
The $\kp_\gm$ excludes the region with $\mch \lsim 200\gev$ and sizable $\lm_{L}$.
The asymmetry of the excluded region by $\kp_\gm$ about $\lm_{L}=0$ 
is attributed to the destructive (constructive) 
interference between $H^\pm$ and $W^\pm$
contributions for $\lm_3>0$ ($\lm_3<0$)~\cite{Arhrib:2015hoa, Ahriche:2018ger}.
The Higgs invisible decay rate limits the value of $\lm_{L}$
very strongly in the mass range of $M_S < \mh/2$.
For $M_S=60\gev$, the maximum allowed value of $|\lm_{L}|$ is only about 0.013, 
which is too small to be seen in this linear scale figure.

\subsection{DM relic density and direct DM detection}
The DM relic density has been measured with high precision
by the PLANCK experiment~\cite{Ade:2015xua}:
\bea
\Omega_{\rm DM}^{\rm Planck} h^2 = 0.1184 \pm 0.0012.
\eea
In the IDM-$U(1)$, both $h_1$ and $h_2$ contribute to the relic density.
With the possibility that there exist other sources of DM,
we avoid the DM overabundance.
At the same time, we do not allow too small 
contribution of our model.
If the relic density of $h_1$ and $h_2$
cannot reach just $1\%$ of the observed relic density,
there must be a more important new physics model providing DM candidates:
the motivation for studying the phenomenology of the IDM-$U(1)$ is getting very weak.
%there must be a more important NP model for the DM:
%the motivation for the serious study on the IDM-$U(1)$ is weak.
Therefore we demand
\bea
\label{eq:relic:density}
0.01  < \frac{\Omega_{h_{1,2}}  }{\Omega_{\rm DM}^{\rm Planck} }  < 1.
\eea

%The direct detection experiments should also be considered.
We also consider the constraints of direct detection experiments. To do so,  we calculate the spin-independent DM-nucleon elastic scattering cross section ($\sigma_{\mathrm{SI}}$) by using \texttt{micrOMEGAs} package~\cite{Belanger:2018mqt}. We then require that $\sigma_{\mathrm{SI}}$ is below the bounds reported on by 
%and compare it with the DM search results from 
the XENON1T experiment~\cite{Aprile:2017iyp}.
In cases that the relic density in our model is smaller than the Planck measurement, we use the rescaled cross section
%Since the inert DM does not explain the whole relic density, we 
\bea
\label{eq:rescaled}
\hat{\sg}_{\rm SI} 
= \frac{\Omega_{h_{1,2}} h^2}{\Omega_{\rm DM}^{\rm Planck} h^2} \, {\sg}_{\rm SI}.
\eea

\begin{figure}[h] \centering
\begin{center}
\includegraphics[width=.8\textwidth]{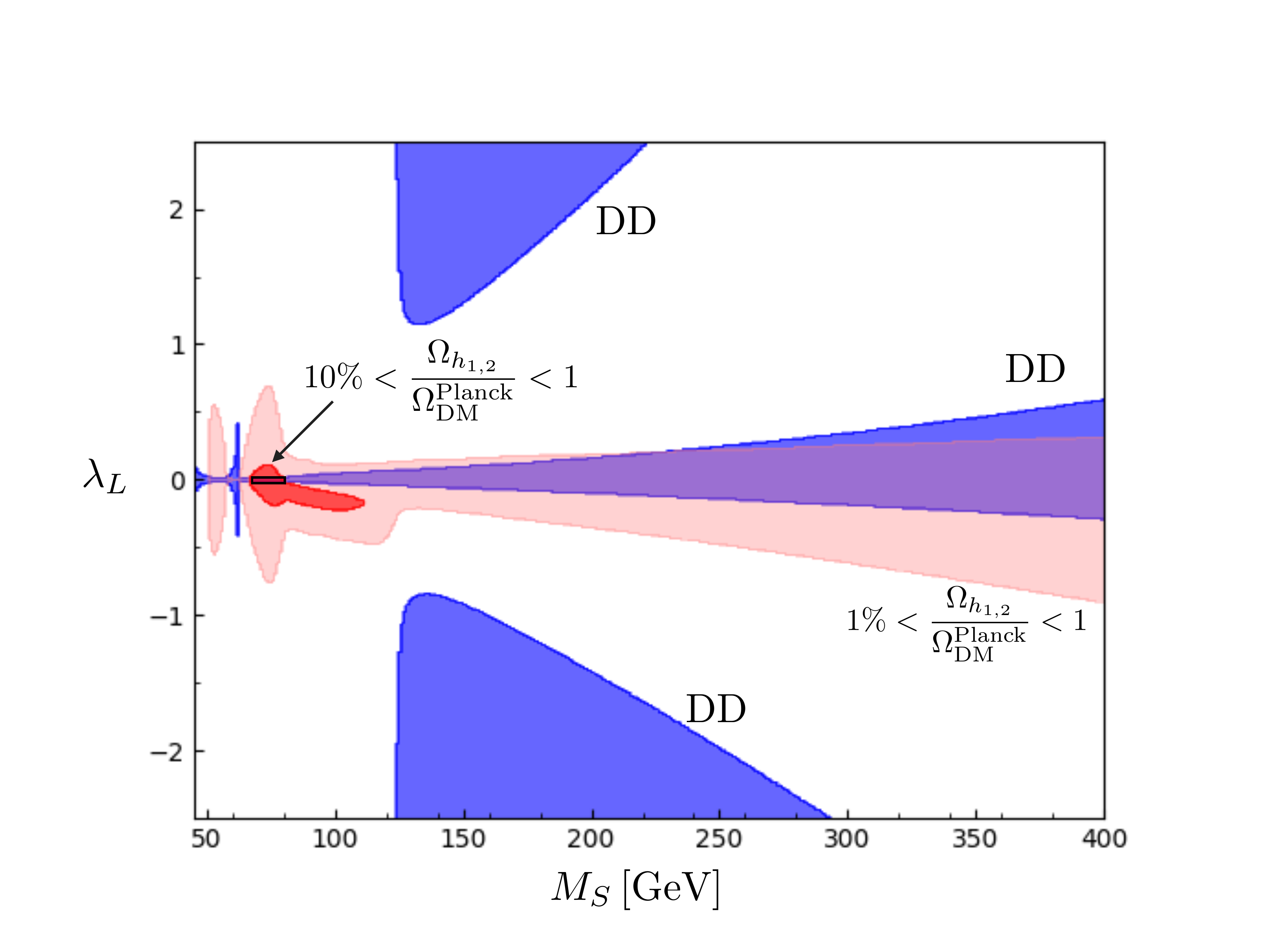}
\caption{\label{fig-relic-DD}
Allowed region of $(M_S,\lm_{L})$ by the relic density
and the direct detection of the DM, by demanding
$
0.01 < \Omega_{h_1,h_2}/\Omega_{\rm DM}^{\rm Planck} <1 $ (pink region), 
$0.1 < \Omega_{h_1,h_2}/\Omega_{\rm DM}^{\rm Planck}<1$ (red 
region),
and the XENON1T experiment (blue region).
We set $\mch = M_S+ 85\gev$.
}
\end{center}
\end{figure}

Figure \ref{fig-relic-DD} shows the allowed region by the relic density
and the XENON1T experiment. 
The pink region is allowed by the condition of 
$0.01 < \Omega_{h_{1,2}}/\Omega_{\rm DM}^{\rm Planck} <1$,
permitting a wide mass range of the inert DM particles as long as $|\lm_L|$ is small enough.
The maximum of $\Omega_{h_{1,2}}/\Omega_{\rm DM}^{\rm Planck}$
is only $\sim 25\%$, occurring at $M_S \simeq 85\gev$ and $\lm_L \simeq -0.13$.
The inert dark scalars alone cannot explain the observed $\Omega_{\rm DM}^{\rm Planck}$.
%We need other sources for the DM.
If we demand $\Omega_{h_{1,2}}/\Omega_{\rm DM}^{\rm Planck} > 10\%$ (red region),
only a small portion of the parameter space around $65 \lsim M_S\lsim 115\gev$
and $ |\lm_L| \ll 1$ survives.
The blue regions are allowed by the XENON1T experiment,
consisting of
the horizontal region with small $|\lm_L|$
and two triangular regions with $|\lm_L|>1$.
The triangular regions are permitted,
because of the suppression from  
very small ${\Omega_{h_{1,2}}  }/{\Omega_{\rm DM}^{\rm Planck} }$
in Eq.~(\ref{eq:rescaled}).
The overlapping region is allowed by the combination of the two constraints.
If we demand
$\Omega_{h_{1,2}}/\Omega_{\rm DM}^{\rm Planck} > 10\%$,
the combined DM constraints exclude most of the parameter space, 
except for very narrow area (red region enclosed by black solid line) 
with $65 \lsim M_S\lsim 80\gev$ and $|\lm_L|<0.01$. 
%\textcolor{green}{(Comment from AJ: what is the value of the relic density for the blue area?)}\\
%\textcolor{green}{\textcolor{green}{Is the allowed region by direct detection allowed by relic density as well? I think only the overlapping regions are allowed by both the constraints.}}.

%%%%%%%%%%%%%%%%%%%%%%%%%%%%%%%
\section{Probing the IDM-$U(1)$ at the LHC}
\label{sec:LHC}
%%%%%%%%%%%%%%%%%%%%%%%%%%%%%%%

\subsection{Production of the inert DM associated with gauge bosons 
at the LHC}

The phenomenology of the IDM-$U(1)$ at the LHC is simple since the model contains only two neutral scalars ($h_1$ and $h_2$), which will play the role of missing energy,
%two missing particles ($h_1$ and $h_2$)
and the charged Higgs boson $H^\pm$ decaying into $W^{\pm (*)} h_{1,2}$. 
The production of the inert scalar bosons is mainly via the gauge bosons since the SM Higgs boson plays a minor role.
The two production channels mediated by the SM Higgs boson,
$gg \to H \to h_i h_i$ and $gg \to H \to H^+ H^-$,
are suppressed:
the vertices $H$-$h_1$-$h_1$ and $H$-$h_2$-$h_2$ are
proportional to the very small $|\lm_L| \lsim \mco(0.01)$;
both channels are one-loop induced with the exchange of an off-shell SM Higgs boson.
In summary, the production of inert scalars are, to a large extent, model-independent due to the sole contribution of gauge couplings. 

For each production channel of the inert scalar bosons,
we attach gauge bosons in order to tag the missing energy signal.\footnote{There are other processes such as the mono-Higgs process and the vector boson fusion production of the inert scalar bosons. However, these processes are sub-leading and we ignore them in this work.}
Limiting up to two gauge bosons as tagging particles, 
the following schematic processes are feasible at the LHC:
\bit
\item $\big[ \qq \to Z^* \to h_1 h_2 \big] \oplus g/\gm/W^\pm/Z$;
\item $\big[ q\bar{q}' \to W^{\pm *} \to H^\pm h_{1,2} \to  W^{\pm (*)} h_{1,2}h_{1,2} \big] \oplus
\gm/Z/W^\pm$;
\item $qq \to Z^*/\gm^{*} \to H^+ H^- \to W^+ W^-  h_{1,2} h_{1,2}$;
\item $\qq \to Z^* \to Z h_i h_i $ and $ q\bar{q}' \to W^{\pm *} \to W^\pm h_i h_i $ ($i=1,2$);
\item $gg \to H \to H^+ H^- \to W^+ W^-  h_{1,2} h_{1,2}$.
\eit
%Here $\oplus V$ denotes tagging the gauge boson $V$ wherever possible.
In terms of final states,
we have mono-jet, mono-$\gm$, mono-$Z$, mono-$W$,
mono-$W\gm$, mono-$WZ$, and mono-$WW$ channels.
As with the terminology of mono-$X$, the mono-$XX'$ process means the production of $XX'$ associated with large missing transverse energy.

\begin{figure}[h] \centering
\begin{center}
\includegraphics[width=.47\textwidth]{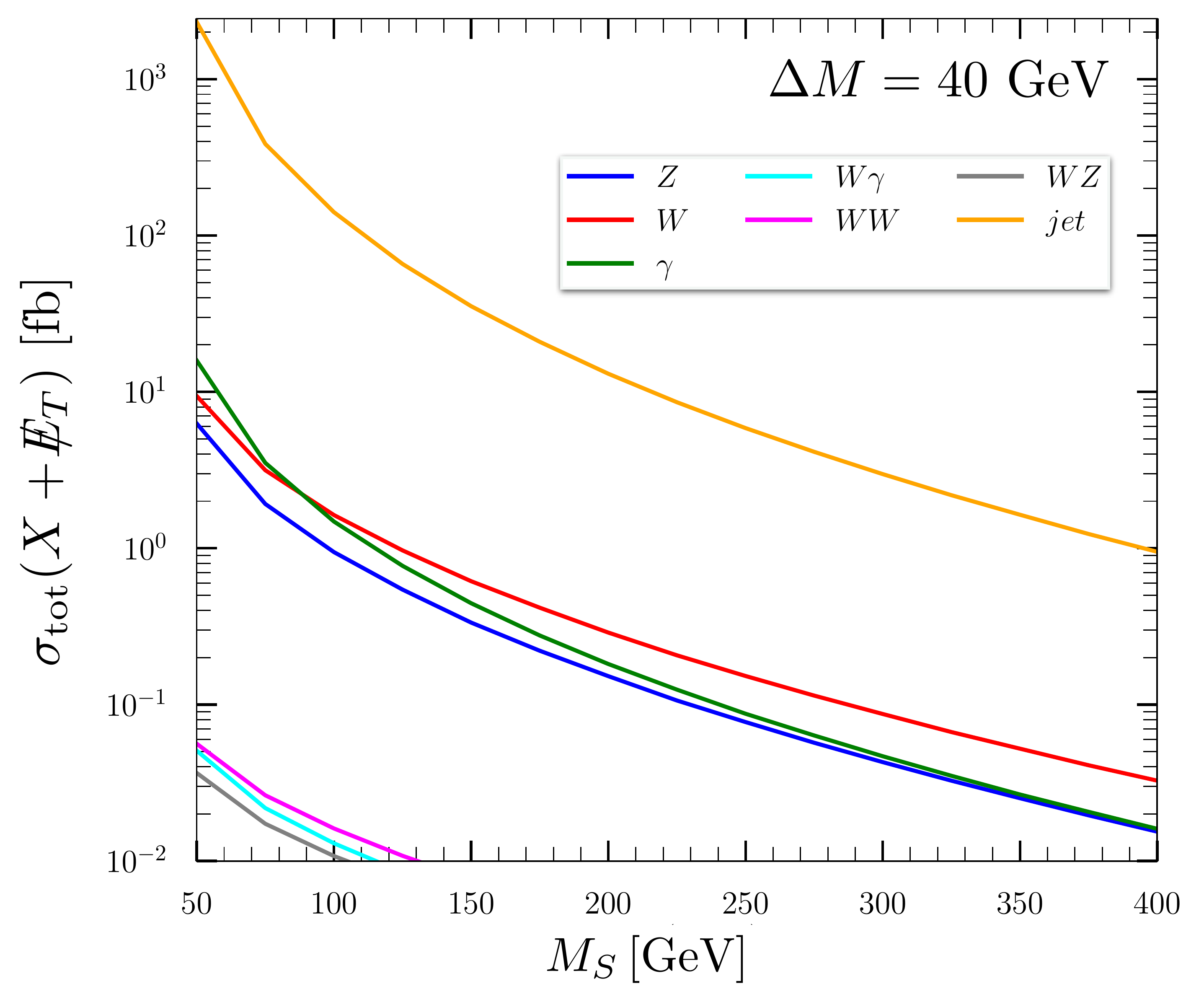}
\includegraphics[width=.47\textwidth]{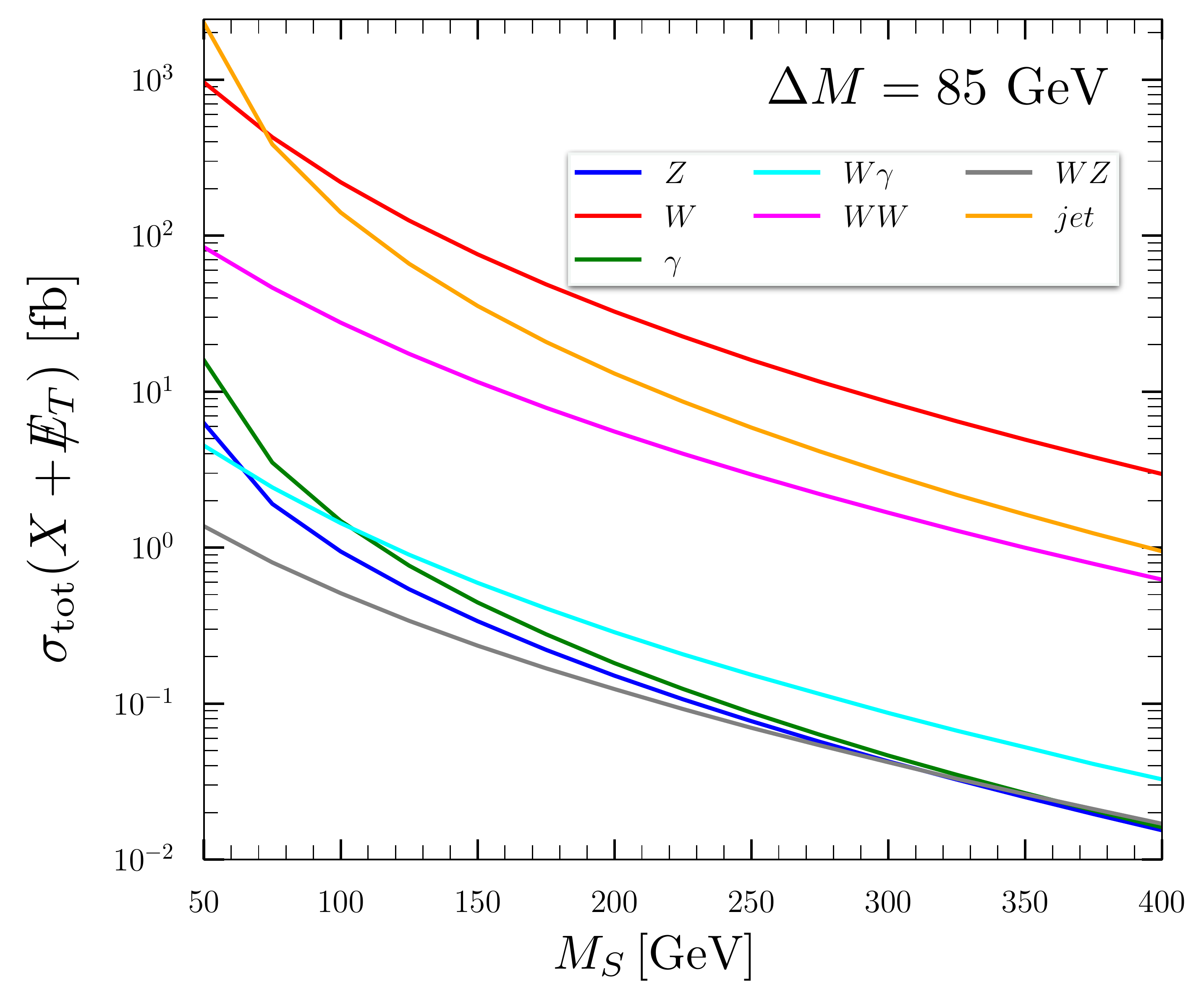}
\caption{\label{fig-production-sigma}
The cross sections of mono-$X$ and mono-$XX'$ processes
at the 14 TeV LHC as a function of $M_S$.
We set $\lm_L=0$, $\lm_2=0.5$,
$\Dt M = \mch-M_S =40\gev$ (left panel), and $\Dt M =85 \gev$ (right panel).
}
\end{center}
\end{figure}

Figure \ref{fig-production-sigma} presents the total production cross sections of various mono-$X$ and mono-$XX'$ processes as a function of $M_S$ at the 14 TeV LHC. 
We have used \textsc{Madgraph5\_aMC@NLO} \cite{Alwall:2011uj, Alwall:2014hca}
with the IDM model file \cite{Goudelis:2013uca, Belanger:2015kga} in the \texttt{Ufo} format \cite{Degrande:2011ua}.
For $\lm_L=0$ and $\lm_2=0.5$, we consider two cases, 
$\Dt M = 40\gev$ (left panel) and $\Dt M = 85\gev$ (right panel). The mono-jet production cross section,
which is independent of $\Dt M$, is the largest,
about 500 fb for $M_S = 70\gev$.
In the $\Dt M  =40\gev$ case, the production cross sections of mono-$V$ and mono-$VV'$,
where $V^{(\prime)}$ is an electroweak gauge boson,
are very small. 
The mono-$V$ processes barely keep 
$\sg_{\mathrm{mono-}V}\sim \mco(1) \fb$
for $M_S \lsim 130\gev$.
The mono-$VV'$ processes have much smaller cross sections, 
below $\sim 0.1 \fb$ even for $M_S = 50\gev$.
%It is very difficult to probe them.

In the $\Dt M  =85\gev$ case,
the production cross sections of the processes involving $W^\pm$ bosons
are highly enhanced thanks to the on-shell decay of $H^\pm\to W^\pm h_{1,2}$.
First, the mono-$W$ cross section becomes comparable to the mono-jet one for $M_S \lsim 100\gev$ and larger for $ M_S \gsim 100\gev$.\footnote{The cross sections of the mono-$W$ process in the $\Dt M  =85\gev$ case are still consistent with the current ATLAS result of the search of DM in association with hadronically-decaying $W/Z$ channel~\cite{Aaboud:2018xdl}.}
The main production channel is
$q\bar{q}' \to  H^\pm (\to W^\pm h_{1,2})  h_{1,2}$, effectively 
a $2 \to 2$ process.
The mono-$WW$ cross section is also enhanced,
as the major production of $q\bar{q}\to H^+(W^+  h_{1,2}h_{1,2}) H^- (\to W^- h_{1,2} h_{1,2})$
is also a $2 \to 2$ process.
The contribution from $gg \to H^* \to H^+ H^-$ is
minor, below $5\%$.
Finally, other interesting processes such as mono-$W\gm$ and mono-$WZ$ yield quite sizable cross sections.

%%%%%%%%%%%%%%%%%%%%%%%%%%%%%%%%%%%%%%%%%%%%%%%%%%%%%%%%%%%
\subsection{Assessing the LHC sensitivities of various mono-$X$ and mono-$XX'$ processes in probing the IDM-$U(1)$ }
%%%%%%%%%%%%%%%%%%%%%%%%%%%%%%%%%%%%%%%%%%%%%%%%%%%%%%%%%%

Before presenting our main results in the next section,
we assess the LHC discovery prospects of some mono-$X$ and mono-$XX'$ processes:
\begin{description}
\item[$\bullet$]\textbf{Mono-jet process}\\ 
In the IDM with $Z_2$ parity, the main production channel for the mono-jet process
is $gg \to H^* \to h_1 h_1 g$. In the IDM-$U(1)$, however, the vertex $H$-$h_i$-$h_i$ ($i=1,2$)
is very suppressed by the combined constraint from the relic density and
the direct detection experiments (see Fig.~\ref{fig-relic-DD}).
Therefore, the main contribution to the mono-jet signature is from $pp \to Z^* \to h_1 h_2 j$.
In Ref.~\cite{Belyaev:2018ext}, this mode was studied as an exceptional case of the IDM-$Z_2$ by imposing the condition of $M_{h_2}-M_{h_1}=1\gev$: 
it was concluded that this process does not reach the discovery at the HL-LHC for $M_{h_1} \gsim 50\gev$. 
It is difficult to probe the IDM-$U(1)$ through the mono-jet channel.
\item[$\bullet$]\textbf{Mono-$W$ process}\\
Even though the on-shell decay of $H^\pm$ into $W^\pm \gm$ helps to increase 
the total production cross section of the mono-$W$ process, 
about $0.5\pb$ when $M_S=70\gev$ and $\Dt M=85\gev$,
it is not easy to probe this mode.
For the hadronic decay of $W^\pm$,
the SM backgrounds such as $pp \to W^\pm Z (\to \bar{\nu}\nu)$ 
and $pp\to jj Z(\to \bar{\nu}\nu)$ are overwhelming.
For the leptonic decay mode of $W^\pm$, yielding the final state of $\ell+\met$,
the irreducible SM background $pp \to W^\pm (\to \ell\nu)$ is enormous:
the observed total cross section at 13 TeV is $\sg(pp \to W^\pm)\simeq 97\nb$~\cite{Aad:2016naf}.
We found that both the signal and the background have very similar shapes 
in the main kinematic distributions
such as the transverse momentum of the charged lepton,
the missing transverse energy $\met$, the total transverse hadronic energy $H_T$,
and the imbalance between the charged lepton and the missing transverse energy 
$p_T^\ell/\vec{E}_T^{\rm miss}$~\cite{Ahriche:2017iar}. The similarity is partially due to light DM mass.
Considering other reducible backgrounds such as $\ttop$,
the Drell-Yan production of dilepton with one lepton escaping the detection,
and the diboson ($WW$, $WZ$, and $ZZ$) productions,
we expect that this mode is challenging to probe at the future LHC.
\item[$\bullet$]\textbf{Mono-$Z$ and mono-$\gm$ process}\\
Both mono-$Z$ and mono-$\gm$ processes
have the production cross sections below $\sim 1 \fb$,
which are almost independent of $\Dt M$.
The cross sections are too small to probe the model.
The current status of the DM searches through mono-$Z$ is 
$\sg_{{\rm mono-}Z} \lsim 3 \pb$,  
at $\sqrt{s}=13\tev$ with the total integrated luminosity of $36.1\ifb$~\cite{Aaboud:2018xdl}. 
The projection to the HL-LHC is 
$\sg_{{\rm mono-}Z} \lsim 300\fb$.
The IDM-$U(1)$ has no chance to be probed through the mono-$Z$ process.
The mono-$\gm$ process is also infeasible to probe in our model~\cite{Aaboud:2017dor,Sirunyan:2018dsf}.
The main reason is that the $\met$ distribution of our signal,
populated below $150 \gev$, is very similar to the SM background.
\item[$\bullet$]\textbf{Mono-$WW$ process}\\ This process yields 
the clean di-lepton plus missing transverse energy signal at the LHC~\cite{Belanger:2015kga,Dolle:2009ft}.
According to the analysis in the IDM with $Z_2$ parity~\cite{Dolle:2009ft},
the signal significance of
this mode is extremely small to be $n_s/\sqrt{n_b}\simeq 0.02$,
for $\mch-M_{h_1}=50\gev$ and $M_{h_2}-M_{h_1}=10\gev$.
Even though our model has a little larger cross section,
it is not enough to enhance the signal-to-background ratio.
\item[$\bullet$]\textbf{Mono-$W \gamma$ process}\\ 
When $\Dt M = 85\gev$, the production cross section of mono-$W\gm$
is about $2.8\fb$ for $M_S=70\gev$.
The ATLAS~\cite{ATLAS:2012gut} and CMS~\cite{Sirunyan:2018psa}
collaborations analyzed this mode
in the search for 
supersymmetry with
a general gauge-mediated mechanism.
As the gravitino $\tilde{G}$ being the lightest supersymmetric particle,
the lightest neutralino $\neuo$ decays into $\gm \tilde{G}$.
The production of $\neuo$ in association with the light chargino $\chao$ 
will yield mono-$W \gm$ signal.
The current 95\% C.L. upper limit on the production cross section is
of the order of $\mco(10) \fb$~\cite{Sirunyan:2018psa}.
We expect higher discovery potential in the future.
We also note that 
this mode has not been studied in the framework of 
the IDM.

\end{description}
Based on these discussions,
we conclude that  the mono-$W\gm$ mode is one of the most sensitive channels
to probe the IDM-$U(1)$ at the LHC.

%%%%%%%%%%%%%%%%%%%%%%%%%%%%%
\section{$ W^\pm \gm  \met $ final states at the LHC}
\label{sec:simulation}
%%%%%%%%%%%%%%%%%%%%%%%%%%%%%

In this section, we make a comprehensive analysis of the mono-$W\gm$ mode
with the hadronic $W^\pm$ decay
at the HL-LHC and a future FCC-hh $100$ TeV collider:  
\bea
pp\to W^\pm(\to  q\bar{q}') \gamma + E_T^{\mathrm{miss}} .
\eea
From the comprehensive study of the theoretical
and experimental constraints on the model, 
we take the following benchmark:
\bea
M_S=70\gev, \quad \mch = 155 \gev, \quad \lm_L=0.01,\quad \lm_2 = 0.5.
\label{eq:benchmark}
\eea
The choice of $\lm_L$ and $ \lm_2$ does not affect the mono-$W\gm$ process. 
The parton-level cross section
of the signal process is
$\sigma\times\mathcal{B}(W^\pm\to q\bar{q}') = 3.23 \fb$ ($29.4 \fb$) 
at $\sqrt{s} = 14\tev~(100\tev)$. 
The final state consists of a hard isolated photon, at least two jets,  
and large missing transverse momentum. 
For this final state, the backgrounds contaminating the searches fall into three categories:  
\begin{itemize}
\item Irreducible backgrounds:
	\bit
	\item $Z(\to \bar{\nu}\nu)\gamma+\mathrm{jets}$;
	\item $Z(\to \bar{\nu}\nu)Z(\to \bar{q}q)\gamma$;
	\item $W^\pm(\to \bar{q}q')  Z(\to \bar{\nu}\nu)\gamma$ .
	\eit
The first background of $Z\gamma+\mathrm{jets}$ is dominant with the total cross section of  $\sim 17 ~(83) \pb$ at the 14 (100) TeV LHC while the other two are sub-leading.
\item Reducible backgrounds involving the leptonic decay of a $W^\pm$ boson
with the charged lepton escaping the detection
(called $\ell^\pm_{\rm esc}$).
We consider three reducible backgrounds:
	\bit
	\item $W^\pm(\to \ell^\pm_{\rm esc}\nu)\gamma+\mathrm{jets}$; 
	\item $W^\pm (\to \ell^\pm_{\rm esc} \nu) W^\mp (\to \bar{q}q') \gamma$
	and $W^\pm(\to\ell^\pm_{\rm esc} \nu) Z(\to q\bar{q}) \gamma$;
    \item $t\bar{t}\gamma$, followed by the semi-leptonic decay of the $t\bar{t}$ pair. 
    \eit 
\item Backgrounds from the fake photons.
There exist non-zero probabilities of misidentifying 
an electron or a jet as a photon.  
The photon fake rates are usually taken as
$P_{j\to \gm}=5\times 10^{-4}$ and $P_{e^- \to \gm}= 2\% ~(5\%)$
in the barrel (endcap) region,
according to the combined study on the perspectives for the HL-LHC
by the ATLAS and CMS collaborations~\cite{Atlas:2019qfx}.
However, they depend sensitively on the type of the process 
as well as the signal region.
For example,
the experimental study on the process $pp \to \gm \met$
in
the signal region with $E_{T}^{\mathrm{miss}} >150\gev$
yields $P_{e^- \to \gm}= 1.5\%$~\cite{Aaboud:2017dor}. 
Since this type of background cannot be modeled in our analysis setup,
especially at the 100 TeV LHC, we will ignore the sub-leading backgrounds from the fake photons.
\end{itemize}

Signal and background processes are simulated at LO, using \texttt{Madgraph5\_aMC@NLO} \cite{Alwall:2011uj, Alwall:2014hca} with the \texttt{NNPDF31} PDF set \cite{AbdulKhalek:2019ihb} and $\alpha_s(M_Z) = 0.118$. 
For the renormalization and factorization scales, we choose 
\beq
\mu_{F,R} = \frac{1}{2} \sum_i \sqrt{p_{T,i}^2 + m_i^2}.
\eeq
For $W^\pm\gamma+\rm jets$ and $Z\gamma+\rm jets$, 
we simulated the productions with jet multiplicity up to two jets 
in the final state and merged them 
according to the \texttt{MLM} merging scheme \cite{Mangano:2006rw} 
with a merging scale $Q_0 = 22.5\gev$. We have confirmed the stability of the calculations with respect to the variation of the merging scale. 

In all of the simulations, we have generated events with some generator-level cuts on the photon transverse momentum $p_T^\gamma > 5$-$10\gev$. 
The decays of  $W$, $Z$, and the top quark were performed 
by using \texttt{MadSpin}~\cite{Artoisenet:2012st}. 
\texttt{Pythia8} was used for the showering and hadronization stages~\cite{Sjostrand:2014zea}. 
To include detector angularity and momentum smearing to the particle-level events, 
we used \texttt{Delphes} as a fast-detector simulation tool~\cite{deFavereau:2013fsa} 
with the templates specific for the HL-LHC and the FCC-hh. 
In the analysis, we cluster jets according to the anti-$k_T$ algorithm~\cite{Cacciari:2008gp} 
with jet radius $D=0.4$ using energy flow as input. 
The clustering of jets was performed by \texttt{Fastjet}~\cite{Cacciari:2011ma}.  

After generating events %with the cut $p_T^\gamma > 5$-$10\gev$,
to be called the ``initial events", 
we take the five \textit{basic} selection steps.
First we demand 
$n_\gamma \geq 1$ and $n_j \geq 2$, i.e.,
at least one photon with $p_T^\gamma > 25\gev$ and $|\eta^\gamma| < 2.47$, 
and at least two jets with $p_T^j > 25 \gev$ and $|\eta^j| < 2.5$. 
The second step is the lepton veto:  
we remove the event which contains at least one isolated lepton (electron or muon) 
with $p_T^\ell > 7\gev$ and $|\eta^\ell| < 2.5$. 
The third one is the $b$-jet veto, 
removing the event if the leading or sub-leading jet is $b$-tagged. 
The fourth and fifth steps are designed to 
reduce the $W^\pm(\to \ell^\pm_{\rm esc}\nu)\gamma+\mathrm{jets}$
and $Z(\to \bar{\nu}\nu)\gamma+\mathrm{jets}$.
We select events that contain one dijet candidate consistent 
with a hadronic decay of $W^\pm$
by demanding two jets to satisfy $\Delta R_{j_1 j_2} < 1$. 
If more than two pairs of jets are satisfying this condition, 
we keep the pair with the minimum $\Delta R_{j_1 j_2}$. 
The dijet with $\Delta R_{j_1 j_2} < 1$
is further required to satisfy 
$|M_{j_1 j_2} - m_W| < 10\gev$ with $m_W = 80.4\gev$. 

\begin{table}[!h]
\begin{center}
  {\renewcommand{\arraystretch}{1.2} 
\begin{tabular}{ c || c | c | c | c | c }
\hline
Selection & Signal & $V\gamma+{\rm jets}$ & $t\bar{t}\gamma$ & $VV\gamma$ & $n_s/\sqrt{n_b}$ \\ 
\hline \hline
Initial events & $9.69\times10^{3}$  & $2.04\times10^8$ & $2.14\times10^6$ & $2.56\times 10^6$ &  $2.54 \times 10^{-3}$ \\ \hline
$n_\gamma \geq 1, n_j \geq 2$ & $1.90\times 10^3$ &  $1.40\times 10^7$ &  $7.60\times 10^5$ &  $4.48\times10^5$ &  $6.83\times 10^{-3}$ \\
Lepton veto & $1.89 \times 10^3$ &   $9.38 \times10^6$ &  $4.02 \times 10^5$ &   $3.72 \times 10^5$ &   $1.02 \times 10^{-2}$ \\
$b$-tag veto & $1.77\times 10^3$ &  $8.94 \times 10^6$ &   $1.40 \times 10^5$ &   $3.30 \times 10^5$ &   $1.03 \times 10^{-2}$ \\
$\Delta R_{j_1 j_2} < 1$ & $5.85 \times 10^2$ &  $1.99 \times 10^6$ &   $8.16 \times 10^5$ & $1.01\times 10^5$ &   $1.48 \times 10^{-2}$ \\
$|M_{j_1 j_2} - m_W| < 10\gev$ & $1.49 \times 10^2$ &  $1.08 \times 10^5$ & $1.30 \times 10^4$ & $2.21\times 10^4$ &  $5.71 \times 10^{-2}$ \\[2pt] \hline
%\hline
%$\dfrac{E_{T}^{\mathrm{miss}}}{\sqrt{H_T}} > 20\sqrt{\mathrm{GeV}}$ &  \multirow{2}{*}{ $19.87$} &  \multirow{2}{*}{$1470$} &   \multirow{2}{*}{$18.18$}  & \multirow{2}{*}{$148.6$} &   \multirow{2}{*}{$0.66$} \\ 
%and $ \met>xx\gev$ & & & & & \\
\cline{1-6}
\end{tabular}
}
\end{center}
\caption{The number of events for the signal and the backgrounds 
after each of the five basic selection steps
at the 14 TeV LHC with the total integrated luminosity $\mathcal{L} = 3~\mathrm{ab}^{-1}$. 
The background $V\gamma+{\rm jets}$ refers to the combination 
of $Z(\to \bar{\nu}\nu)\gamma+\mathrm{jets}$
and $W^\pm(\to \ell^\pm_{\rm esc}\nu)\gamma+\mathrm{jets}$,
where $\ell^\pm_{\rm esc}$ denotes a charged lepton escaping the detection.
The background $VV\gamma$ includes
$Z(\to \bar{\nu}\nu)Z(\to \bar{q}q)\gamma$, $W^\pm(\to \bar{q}q')  Z(\to \bar{\nu}\nu)\gamma$,
$W^\pm (\to \ell^\pm_{\rm esc} \nu) W^\mp (\to \bar{q}q') \gamma$,
and $W^\pm (\to \ell^\pm_{\rm esc} \nu) Z (\to \bar{q}q) \gamma$.}
\label{tab:cutflow14}
\end{table}

\begin{table}[!h]
\begin{center}
\begin{tabular}{ c | c | c | c | c | c }
\cline{1-6}
Selection & Signal & $V\gamma+{\rm jets}$ & $t\bar{t}\gamma$ & $VV\gamma$ & $n_s/\sqrt{n_b}$ \\ 
\hline\hline 
Initial events                               & $8.82\times 10^5$  & $1.23\times10^{10}$ & $8.70\times10^8$ & $1.76\times 10^8$ &  $1.15 \times 10^{-2}$ \\ \hline
$n_\gamma \geq 1, n_j \geq 2$  & $1.71\times10^5$ &  $1.46\times 10^9$ &  $3.36\times 10^8$ &  $3.13 \times10^7$ &  $1.62\times 10^{-2}$ \\
Lepton veto                                 & $1.71 \times 10^5$ &   $8.73 \times10^8$ &  $1.67 \times 10^8$ &   $2.55 \times 10^7$ &   $2.78 \times 10^{-2}$ \\
$b$-tag veto                               & $1.64\times 10^5$ &  $8.33 \times 10^8$ &   $5.65 \times 10^7$ &   $2.30 \times 10^7$ &   $3.11 \times 10^{-2}$ \\
$\Delta R_{j_1 j_2} < 1$ & $6.23 \times 10^4$ & $2.09 \times 10^8$  &   $3.38 \times 10^7$ & $7.97\times 10^6$ &   $4.31 \times 10^{-2}$ \\
$|M_{j_1 j_2} - m_W| < 10\gev$  & $1.86\times 10^4$ & $1.66 \times 10^7$ & $6.64 \times 10^6$ & $2.02\times 10^6$ &  $1.28 \times 10^{-1}$ \\[2pt] \hline
%\hline
%$\dfrac{ E_{T}^{\mathrm{miss}} }{ \sqrt{H_T} } > 15 \sqrt{\mathrm{GeV}}$ & $9.60\times 10^3$ & $1.35\times 10^6$ & $1.43\times 10^5$ & $8.47 \times 10^4$ & $7.64$ \\
\cline{1-6}
\end{tabular}
\end{center}
\caption{Same as Table \ref{tab:cutflow14} but for the FCC-hh at $\sqrt{s}=100$ TeV and $\mathcal{L} = 30~\mathrm{ab}^{-1}$.}
\label{tab:cutflow100}
\end{table}

In Tables \ref{tab:cutflow14} and \ref{tab:cutflow100},
we show the cut-flows for the signal and the backgrounds at the 14 TeV and 100 TeV LHC, respectively.
In order to assess the discovery potential of the signal process, 
we present the signal significance at each selection step,
defined by~\cite{Cowan:2010js}
\begin{eqnarray}
\mathcal{S} = \frac{n_s}{\sqrt{n_b}},
\end{eqnarray}
where $n_s$ and $n_b$ are the number of events for the signal and backgrounds, respectively.
Among the basic selection steps, 
the $W$-boson mass requirement significantly reduces
the expected number of background events.
However, the backgrounds are still overwhelming,
yielding the significance of the order of $10^{-2}$ ($10^{-1}$)
at the 14 TeV (100 TeV). Therefore, we need to devise a new method in order to enhance the
significance.
%We need to devise a new method.

\begin{figure}[!tbp]
\centering
\includegraphics[width=0.49\linewidth]{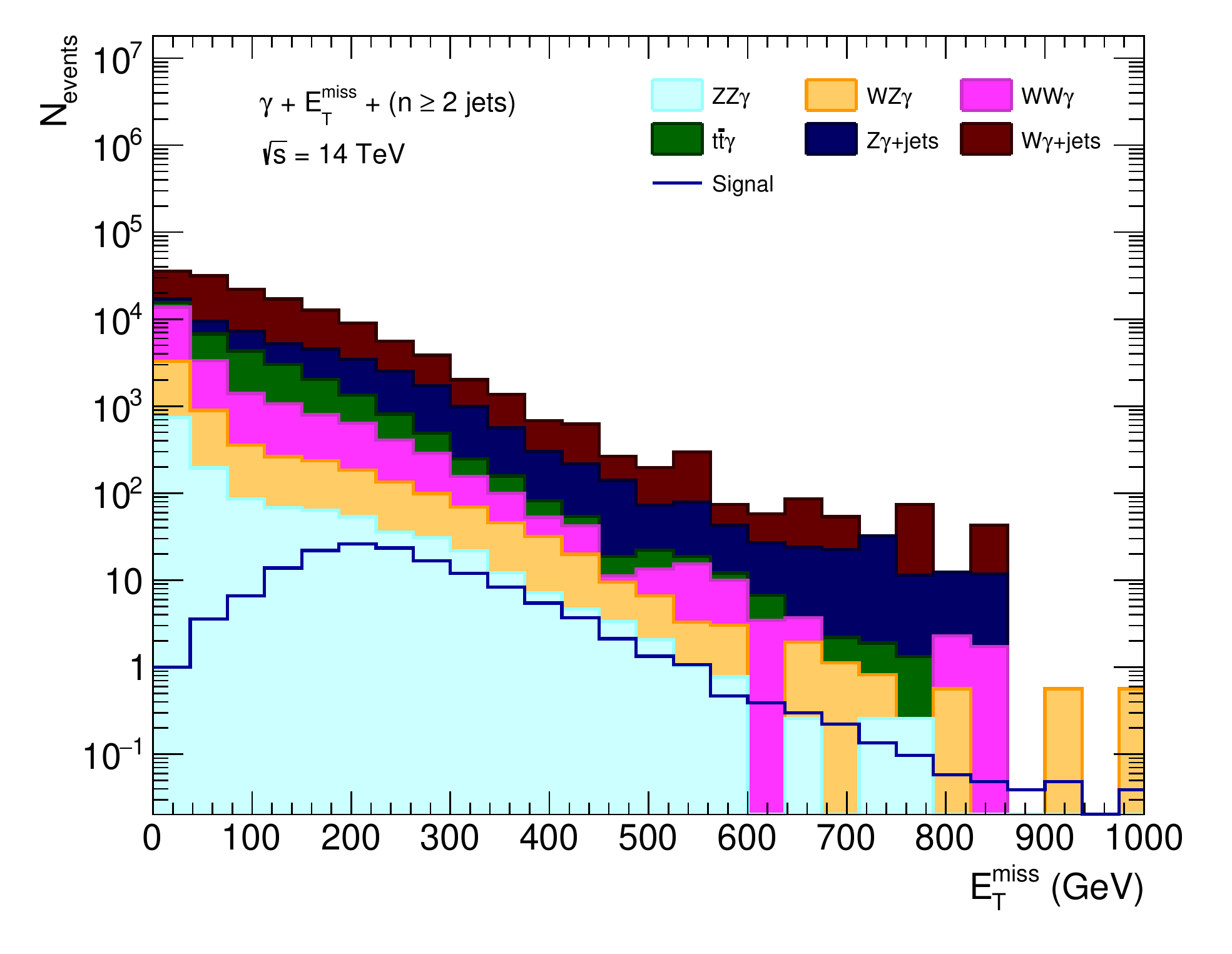}
\hfill
\includegraphics[width=0.49\linewidth]{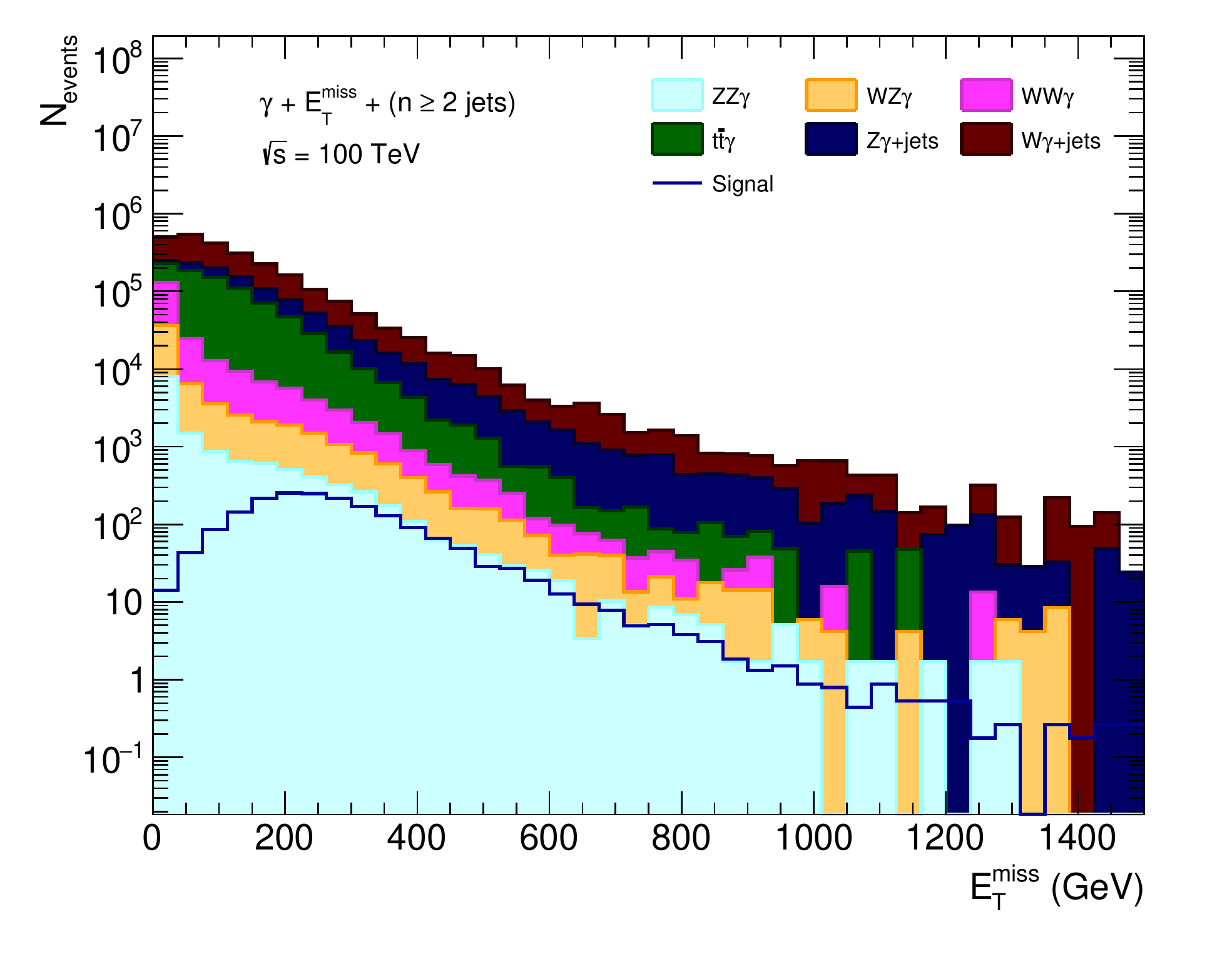}
\caption{Expected number of events for the signal (blue line) 
and the backgrounds stacked on the top of each other as a function of the missing transverse energy at the HL-LHC (left panel) and the FCC-hh (right panel). The distributions are shown after the basic selections.}
\label{fig:MET}
\end{figure}

\begin{figure}[!tbp]
\includegraphics[width=0.49\linewidth]{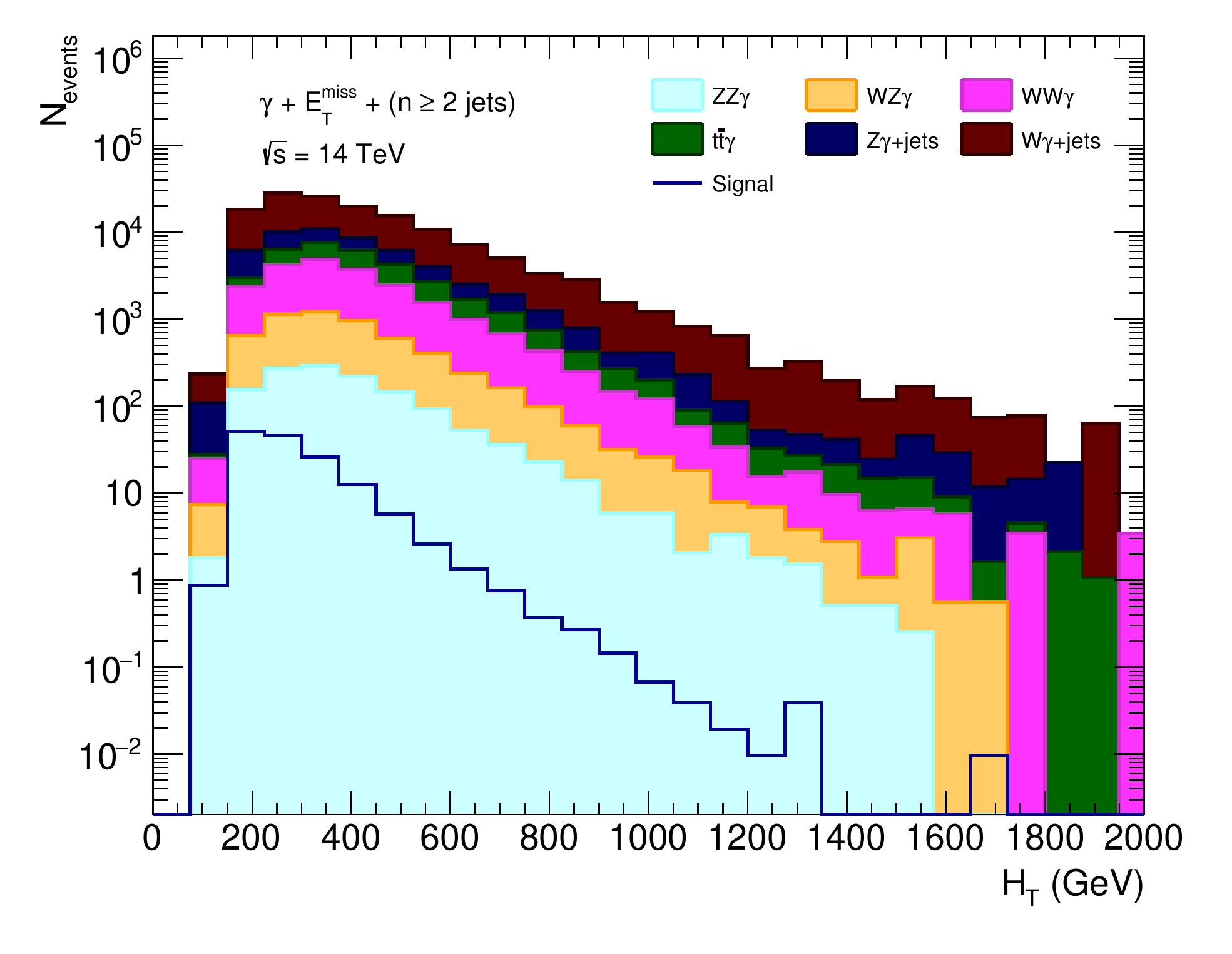}
\hfill
\includegraphics[width=0.49\linewidth]{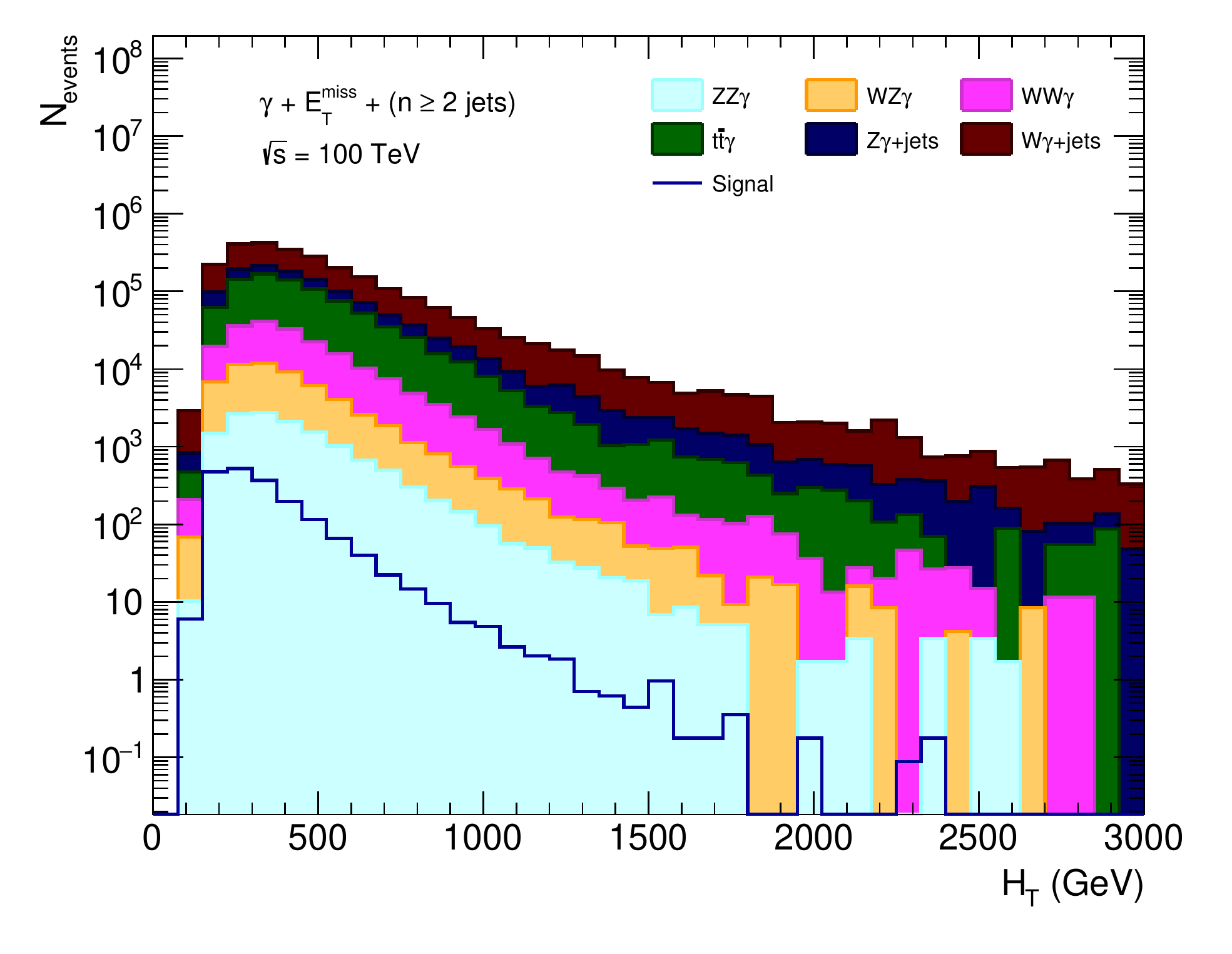}
\caption{Same as Fig.~\ref{fig:MET} but for the scalar sum of the transverse momenta of jets.}
\label{fig:HT}
\end{figure}

\begin{figure}[!tbp]
\centering
\includegraphics[width=0.49\linewidth]{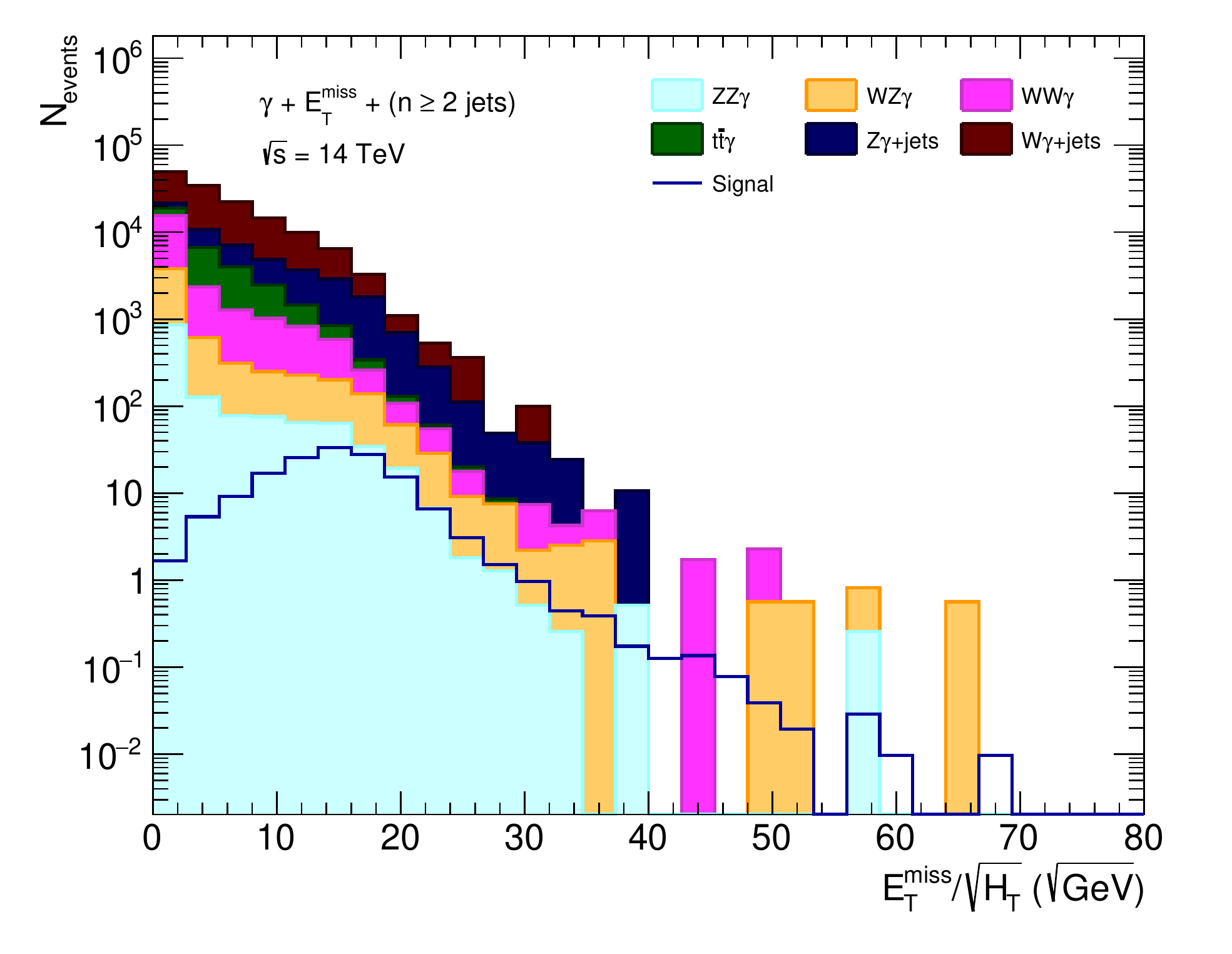}
\hfill
\includegraphics[width=0.49\linewidth]{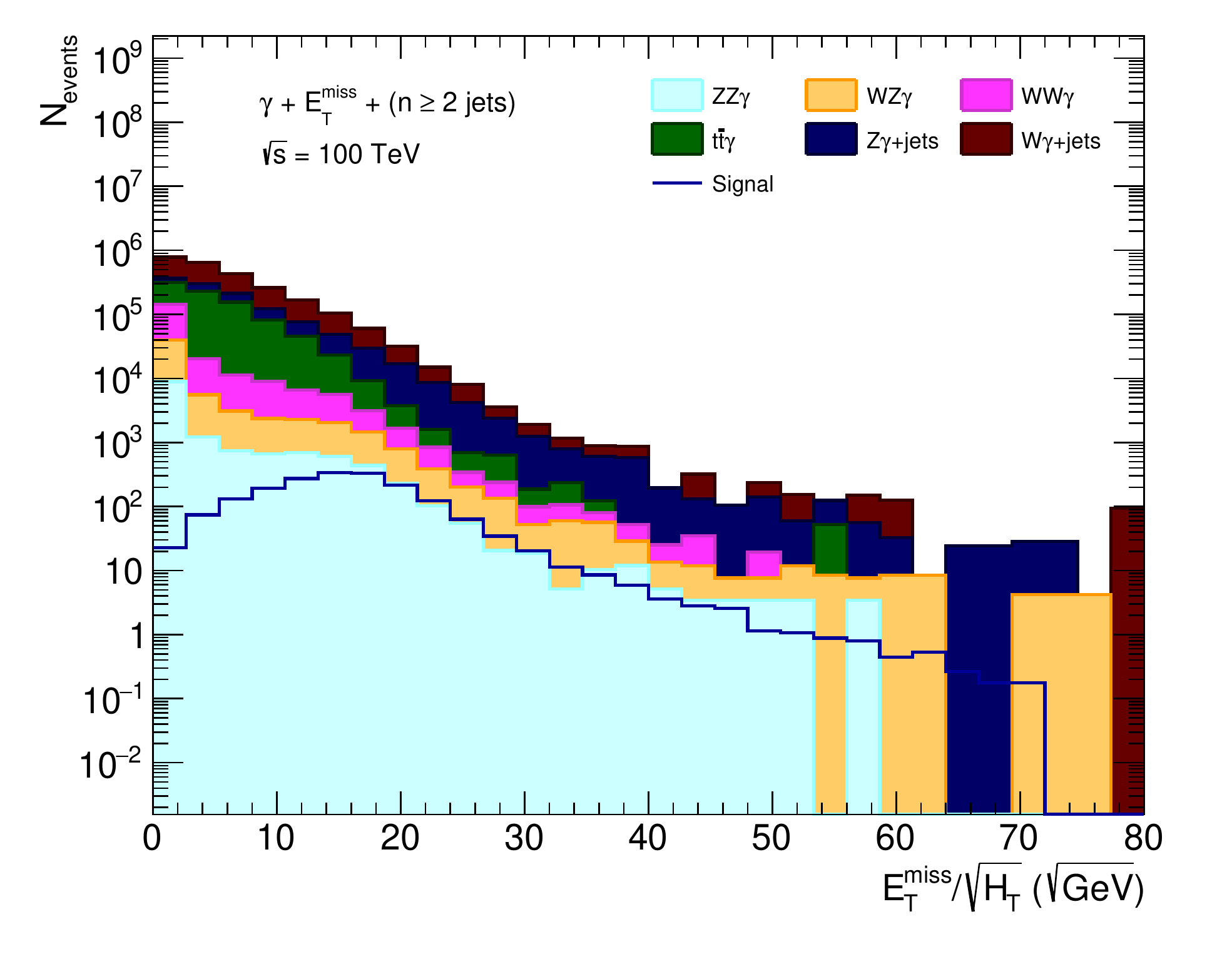}
\caption{Same as Fig. \ref{fig:MET} but for the modified $\met$ significance
${E_T^\mathrm{miss}}/{\sqrt{H_T}}$. }
\label{fig:metHT}
\end{figure}

In Figs.~\ref{fig:MET} and \ref{fig:HT},
we present the distributions of the missing transverse energy,
$\met$, and the scalar sum of
the transverse momenta of jets, $H_T$,
 at the HL-LHC (left panel) and the FCC-hh (right panel).
Unfortunately, the $\met$ distributions of the signal 
have a similar distribution shape to those of the backgrounds. 
Requiring large missing transverse energy does not improve the signal-to-background ratios.
This is attributed to the light DM mass, $M_S=70\gev$,
%of the benchmark point in Eq.~(\ref{eq:benchmark})
which is inevitable for the condition $\Omega_{h_{1,2}}/\Omega_{\rm DM}^{\rm Planck} >0.1$ to be satisfied.
The $H_T$ distribution shows slight differences
between the signal and the backgrounds
such that 
the backgrounds have stronger $H_T$ than the signal.
Based on these characteristics,
we suggest the use of 
the modified $\met$ significance, $\mathcal{P}_{E_T^{\mathrm{miss}}}$, 
defined by 
\begin{equation}      
\mathcal{P}_{E_T^{\mathrm{miss}}} = \frac{E_T^\mathrm{miss}}{\sqrt{H_T}}.
\label{eq:purity}  
\end{equation}
The event yields for $\mathcal{P}_{E_T^{\mathrm{miss}}}$ 
at the HL-LHC and the FCC-hh are shown 
in Fig.~\ref{fig:metHT}. 
We can see that the signal has a peak around 
$\mathcal{P}_{E_T^{\mathrm{miss}}} \simeq 15\;\sqrt{\mathrm{GeV}}$ 
while all of the backgrounds 
have monotonically decreasing shape.
Therefore, selecting $\mathcal{P}_{E_T^{\mathrm{miss}}}$
larger than a certain value, called $\mathcal{C}_{\rm min}$,
could help to improve the signal significance.

\begin{figure}[!tbp]
\includegraphics[width=0.49\textwidth]{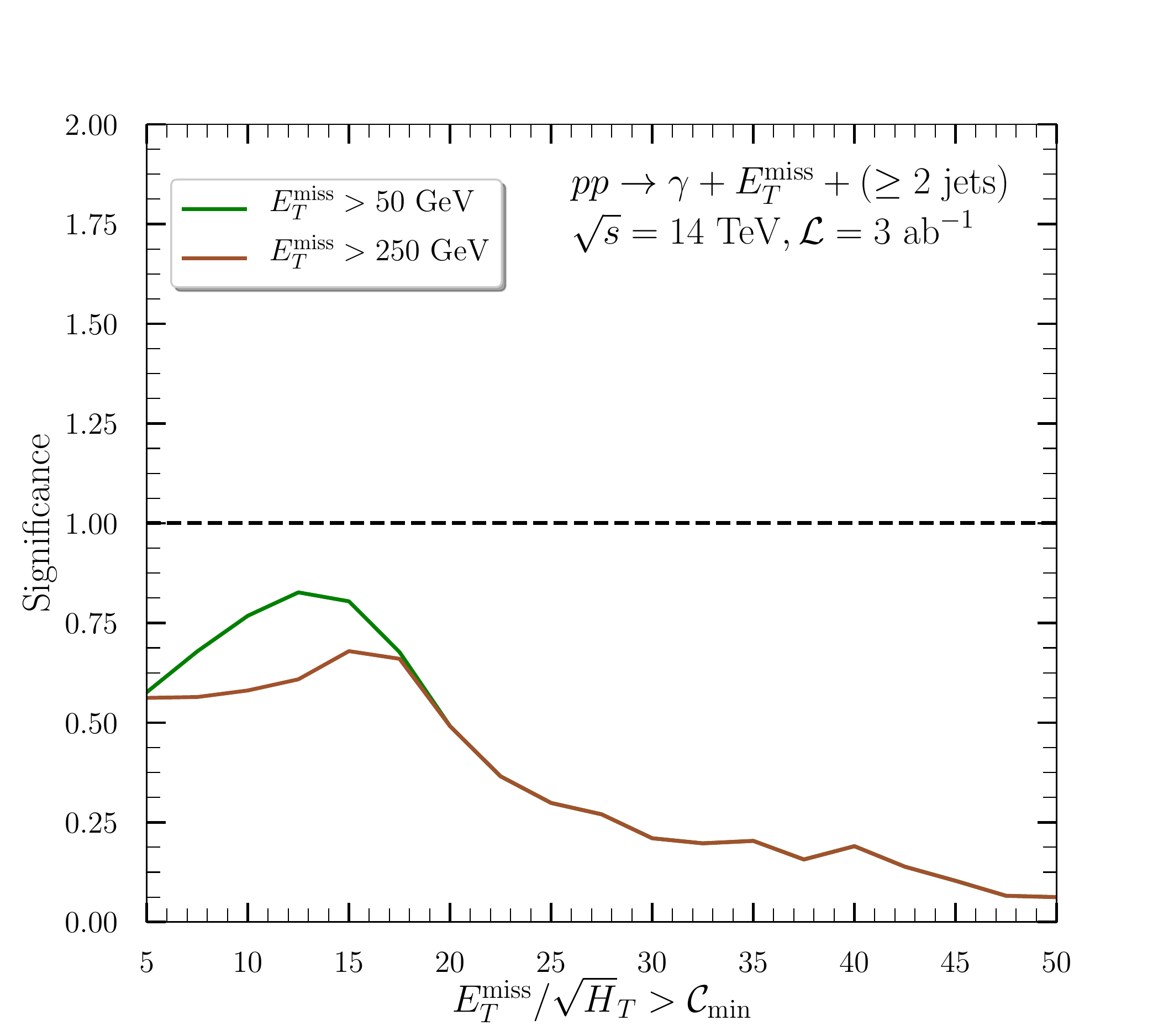}
\includegraphics[width=0.49\textwidth]{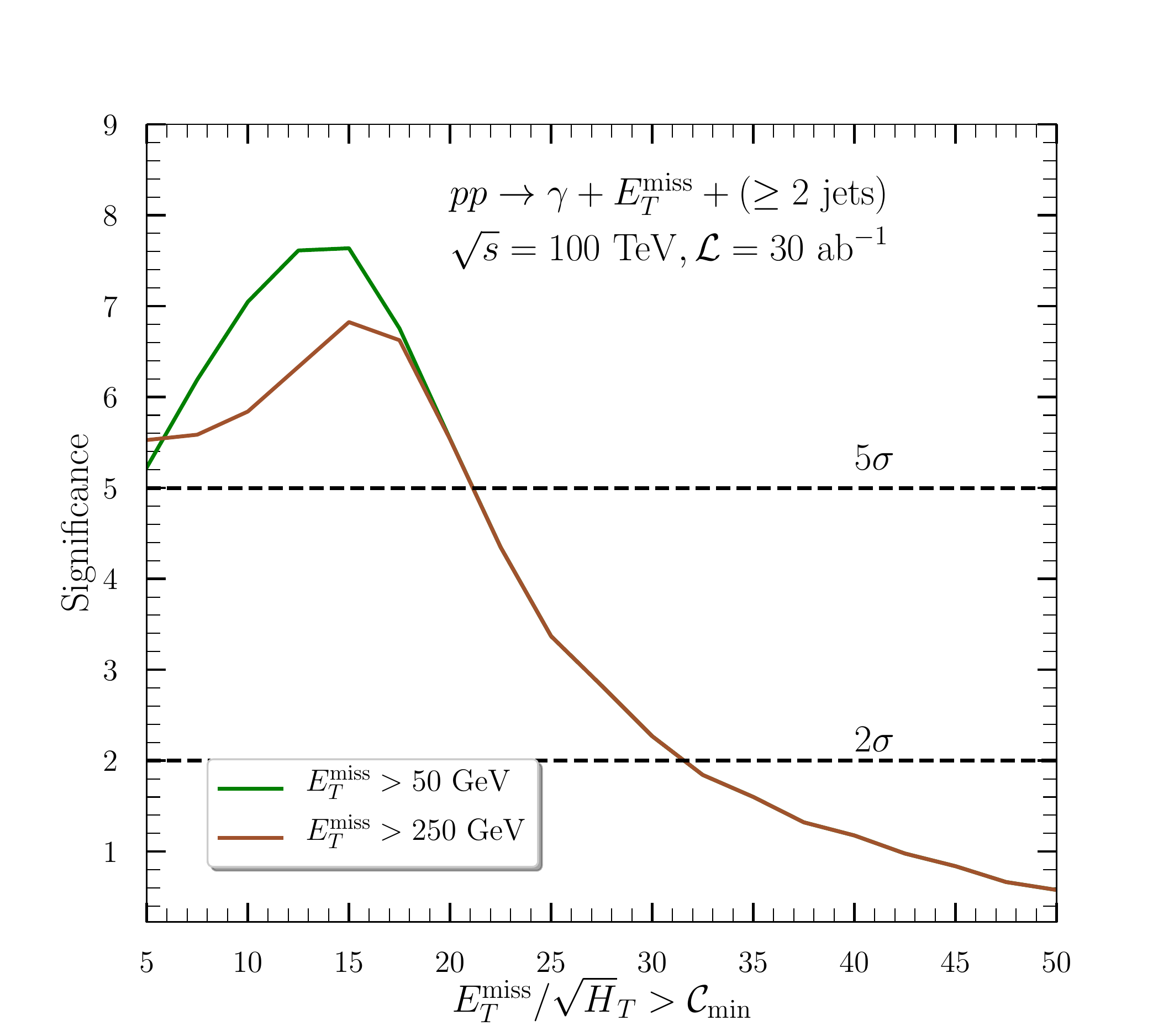}
\caption{Signal significance as a function of the cut on the modified $\met$ significance 
$\mathcal{P}_{E_T^\mathrm{miss}}$ at the HL-LHC  and the FCC-hh  (right panel).}
\label{fig:significance}
\end{figure}

In order to find the optimized cut on $\mathcal{P}_{E_T^{\mathrm{miss}}}$,
we present the significance as a function of 
$\mathcal{C}_{\rm min}$
for the HL-LHC (left panel) and for the FCC-hh at $100$ TeV (right panel)
in Fig. \ref{fig:significance}. 
We consider two cases of $\met>50\gev$ (green line) and $\met>250\gev$ (brown line).
Both at the HL-LHC and FCC-hh,
the cut of $\met>50\gev$ yields higher significance than the cut of $\met>250\gev$
for $ \mathcal{C}_{\rm min} \lsim 18\gev$.
Therefore, the optimal cuts on $\met$ and $\mathcal{P}_{E_T^{\mathrm{miss}}}$ are
\bea
\met \geq 50\gev, \quad \mathcal{P}_{E_T^{\mathrm{miss}}} \geq 15 \sqrt{\mathrm{GeV}}.
\eea
Then the signal significance at the LH-LHC is about $0.77$,
leading to a disappointing result that 
the IDM with $U(1)$ symmetry cannot be probed at the HL-LHC.
At the FCC-hh with $\sqrt{s}=100\tev$ and $30\iab$, 
the significance can be as high as about $7.5$.
The IDM-$U(1)$ has a chance to be probed through the mono-$W\gm$ mode. 

\section{Conclusions}
\label{sec:conclusion} 
We have comprehensively studied the phenomenology of the inert doublet model (IDM)
with a global $U(1)$ symmetry.
The model has an additional Higgs doublet field $\Phi'$,
accommodating new scalar bosons, $h_1$, $h_2$, and $H^\pm$.
By promoting the usually adopted $Z_2$ parity into a global $U(1)$ symmetry,
the theory 
provides two DM particles, neutral \textit{CP}-even and \textit{CP}-odd scalar bosons,
$h_1$ and $h_2$.
The extended symmetry strongly constrains the model.
As the $\lm_5$ term in the scalar potential is prohibited by the $U(1)$ symmetry,
two important results occur.
First, $h_1$ and $h_2$ have the same mass and thus become DM candidates.
Second, the number of model parameters is 
one less than that in the IDM with $Z_2$ parity.
The theory becomes very limited by the combination of various constraints
such as the LEP experiments, electroweak oblique parameters,
theoretical stability, Higgs precision data, DM relic density,
and DM direct detection experiments.
Particularly when we demand that the IDM-$U(1)$
explains at least 10\% of the observed DM relic density,
the parameter space that survived is very narrow:
the DM mass $M_S$ is about $70\gev$;
the parameter $\lm_L$, which governs the couplings of $h_1$ and $h_2$
to the SM-like Higgs boson $H$,
is extremely small;
the charged Higgs boson cannot be heavier than about $M_S+100\gev$.

For the LHC phenomenology of the model,
we first studied the production cross sections of the inert scalar bosons
in the mono-$X$ and mono-$XX'$ channels.
The key factor is the mass difference between the charged Higgs boson
and the DM particles, $\Dt M$.
If $\Dt M$ is above the $W$ boson mass so that the charged Higgs boson
decays into $W^\pm h_{1,2}$ on-shell,
the production cross sections involving $W^\pm$, mono-$W$, mono-$WZ$, mono-$WW$, 
and mono-$WW$, are highly enhanced, compared with the case of $\Dt M < m_W$.
Nevertheless, the cross sections are still small.
Based on the projection of the current 
experimental results of the $13$ TeV LHC into the HL-LHC,
we expected that the mono-$W\gm$ process has a high chance to probe the model.

Focusing on the hadronic decay mode of the $W^\pm$ boson,
we have completed the analysis with full detector-level simulations
for the mono-$W\gm$ signal.
%The backgrounds from
%$W\gm$+jets, $Z\gm$+jets, $\ttop\gm$, and $VV\gm$ were analyzed.
Partially because the DM mass is rather light like $\sim 70\gev$ in this model,
the kinematical distributions of the transverse momentum of the leading-isolated photon and the missing transverse energy are similar for both the signal and the backgrounds.
We found that the scalar sum of the transverse momenta of jets, $H_T$, for the signal
is softer than that for the backgrounds.
So we suggested the use of the modified $\met$ significance, 
$\mathcal{P}_{E_T^{\mathrm{miss}}} =\met/\sqrt{H_T}$,
as a key observable to reduce the backgrounds:
requiring $\mathcal{P}_{E_T^{\mathrm{miss}}}>\mathcal{C}_{\rm min}$ 
is shown to be very efficient.
Although the cut on $\mathcal{P}_{E_T^{\mathrm{miss}}}$ enhances 
the signal significance about tenfold,
the maximum significance at the 14 TeV LHC
with the total integrated luminosity of $3\iab$ is below one.
It is not feasible to probe the model at the HL-LHC.
Extending the analysis into the future FCC-hh at $\sqrt{s}=100\tev$
with the total integrated luminosity of $30\iab$,
the signal significance is shown to reach about 7.5.
The IDM-$U(1)$ has a chance to be probed through the mono-$W\gm$ mode.

\acknowledgments
This work is supported by the National Research Foundation of Korea, Grant No. NRF-2019R1A2C1009419.
The work of S.S. is supported by IBS under the project code, IBS-R018-D1.

%\bibliography{0literature}
%\bibliographystyle{JHEP}

\end{document}